\documentclass[aps , pra , twocolumn , showpacs , superscriptaddress , amsmath , amssymb , reprint]{revtex4-2}

\pdfoutput = 1

\usepackage{graphicx}
\usepackage{epstopdf}
\usepackage{amsmath}
\usepackage{amsfonts}
\usepackage{amssymb}
\usepackage{xcolor, soul}
\sethlcolor{green}
\usepackage[normalem]{ulem}
\setstcolor{red}
\usepackage{siunitx}
\usepackage{bm}
\usepackage[utf8]{inputenc}
\usepackage[T1]{fontenc}
\usepackage{color}
\definecolor{blue(pigment)}{rgb}{0.15, 0.15, 0.7}
\usepackage{url}
\usepackage[hyperfootnotes=false,hypertexnames=false]{hyperref}
\hypersetup
{
	pdftitle={In-plane optical phonon modes of current-carrying graphene},
	bookmarksnumbered = true,
	bookmarksopen = true,
	bookmarksopenlevel = 1,
	colorlinks = true,
	citecolor = blue(pigment),
	linkcolor = blue(pigment),
	urlcolor = blue(pigment)
}

\newcommand{\0}{\textsc{\tiny{(0)}}}

\newcommand{\ket}[1]{\ensuremath{\left|#1\right\rangle}}

\begin{document}

\title{In-plane optical phonon modes of current-carrying graphene}

\author{Mohsen Sabbaghi}
\email{sabbagh2@uwm.edu}
\affiliation{Department of Electrical Engineering, University of Wisconsin-Milwaukee, Milwaukee, WI 53211, USA}

\author{Tobias Stauber}
\email{tobias.stauber@csic.es}
\affiliation{Materials Science Factory, Instituto de Ciencia de Materiales de Madrid, CSIC, E-28049 Madrid, Spain}
\affiliation{Theoretical Physics III, Center for Electronic Correlations and Magnetism, Institute of Physics, University of Augsburg, D-86135 
Augsburg, Germany}

\author{Hyun-Woo Lee}
\email{hwl@postech.ac.kr}
\affiliation{Department of Physics, Pohang University of Science and Technology, Pohang 37673, Korea}

\author{J. Sebastian Gomez-Diaz}
\email{jsgomez@ucdavis.edu}
\affiliation{Department of Electrical and Computer Engineering, University of California-Davis, Davis, CA 95616, USA}

\author{George W. Hanson}
\email{george@uwm.edu}
\affiliation{Department of Electrical Engineering, University of Wisconsin-Milwaukee, Milwaukee, WI 53211, USA}

\date{\today}

\begin{abstract}
In this work, we study the in-plane optical phonon modes of current-carrying single-layer graphene whose coupling to the $\pi$ electron gas is strong. Such modes are expected to undergo a frequency shift compared to the non-current-carrying state due to the non-equilibrium occupation of the Dirac cone electronic eigen-states with the flowing $\pi$ electron gas. Large electron-phonon coupling (EPC) can be identified by an abrupt change in the slope of the phonon mode dispersion known as the Kohn anomaly, which mainly occurs for (i) the in-plane longitudinal/transverse optical (LO/TO) modes at the Brillouin zone (BZ) center ($\Gamma$ point), and (ii) the TO modes at the BZ corners ($K$ points). We show that the breaking of the rotational symmetry by the DC current results in different frequency shifts to the $\Gamma$-TO and $\Gamma$-LO modes. More specifically, the DC current breaks the TO-LO mode degeneracy at the $\Gamma$ point which ideally would be manifested as the splitting of the Raman G peak.
\end{abstract}


\maketitle
%%%%%%%%%%%%%%%%%%%%%%%%%%%%%%%%%%%%%%%%%%%%%%%%%%%%%%%%%%%%%
%SECTION I: INTRODUCTION
%%%%%%%%%%%%%%%%%%%%%%%%%%%%%%%%%%%%%%%%%%%%%%%%%%%%%%%%%%%%%
\section{Introduction}\label{sec:Intro}

Recently, growing interest in studying the impact of DC current on surface plasmon polaritons (SPPs) in graphene \footnote{Each of the one-atom-thick layers of crystalline graphite when isolated from the other layers is referred to as graphene \cite{Piscanec_2004,Malard_2009}.} has emerged on both theoretical \cite{Strikha,ZhaoPeet,Sabstauber_2015,Mikhailov_2016,Duppen_2016,Sabstauber_2018,Wegner_2018,Svintsov_2018,Svintsov_2019,Serrano_2019,Hassani_2022,Sammon_2021} and experimental \cite{Dong_2021,Zhao_2021} fronts. The current-driven drag of SPPs can be described by the non-equilibrium (NE) electromagnetic response of the flowing $\pi$ electron gas \cite{Strikha,ZhaoPeet,Sabstauber_2015,Mikhailov_2016,Duppen_2016,Sabstauber_2018}.

The flow of the $\pi$ electron gas also alters the static dielectric screening properties \cite{Sabstauber_2015}; therefore, the screening of the ``electrostatic'' interaction among the positively-charged carbon ions is altered by DC current. This translates into the modification of the ``spring constant'' of the carbon pairs due to their immersion in the flowing $\pi$ electron gas. Ultimately, this hints at the possibility that the phonon mode frequencies in graphene could be impacted by DC electric current.

The phononic dispersion of graphene can be calculated by constructing the dynamical matrix based on a purely ionic potential  \cite{Jishi_1982,Ferrari_2000,Wirtz_2004,Pisana,Falkovsky_2007,Falkovsky_2008} which leads to the \textit{bare} phonon frequencies. However, the screening of the inter-ionic Coloumb interaction due to the $\pi$ electron gas must be taken into account which leads to the renormalization of the bare phonon frequencies \cite{Piscanec_2004,Lazzeri_2006,Lazzeri_2008}. Such renormalization is the main mechanism through which certain phonon modes are impacted by DC current, and the proposed effect will be discussed in detail in this paper.

Graphene is described by a two-atomic unit cell repeating in two dimensions \cite{NETO_2009}, which leads to the emergence of $6$ phonon modes corresponding to the $6$ degrees of freedom of this $2$-atom building block \cite{Wirtz_2004}. We represent these modes by $\left(\nu , \bm{q}\right)$ with $\nu$ and $\bm{q}$ being respectively the branch index and the momentum vector of the phonon mode. Formally, these phonon modes can be obtained by solving the $\pmb{\mathrm{D}}\!\left(\bm{q}\right) \cdot \hat{\bm{\mathrm{e}}}_{\nu,\bm{q}} = \omega^{2}_{\nu, \bm{q}} \hat{\bm{\mathrm{e}}}_{\nu,\bm{q}}$ eigen-value equation, with $\pmb{\mathrm{D}}\!\left(\bm{q}\right) $, $\hat{\bm{\mathrm{e}}}_{\nu,\bm{q}}$ and $ \omega_{\nu, \bm{q}}$ being respectively the $6\times 6$ dynamical matrix, the $6$-dimensional mode eigen-vector and the mode eigen-frequency, i.e., mode frequency \cite{Michel_2008,Stauber_2008}.

\begin{figure}[t!]
	\begin{center}
		\includegraphics[width = \columnwidth]{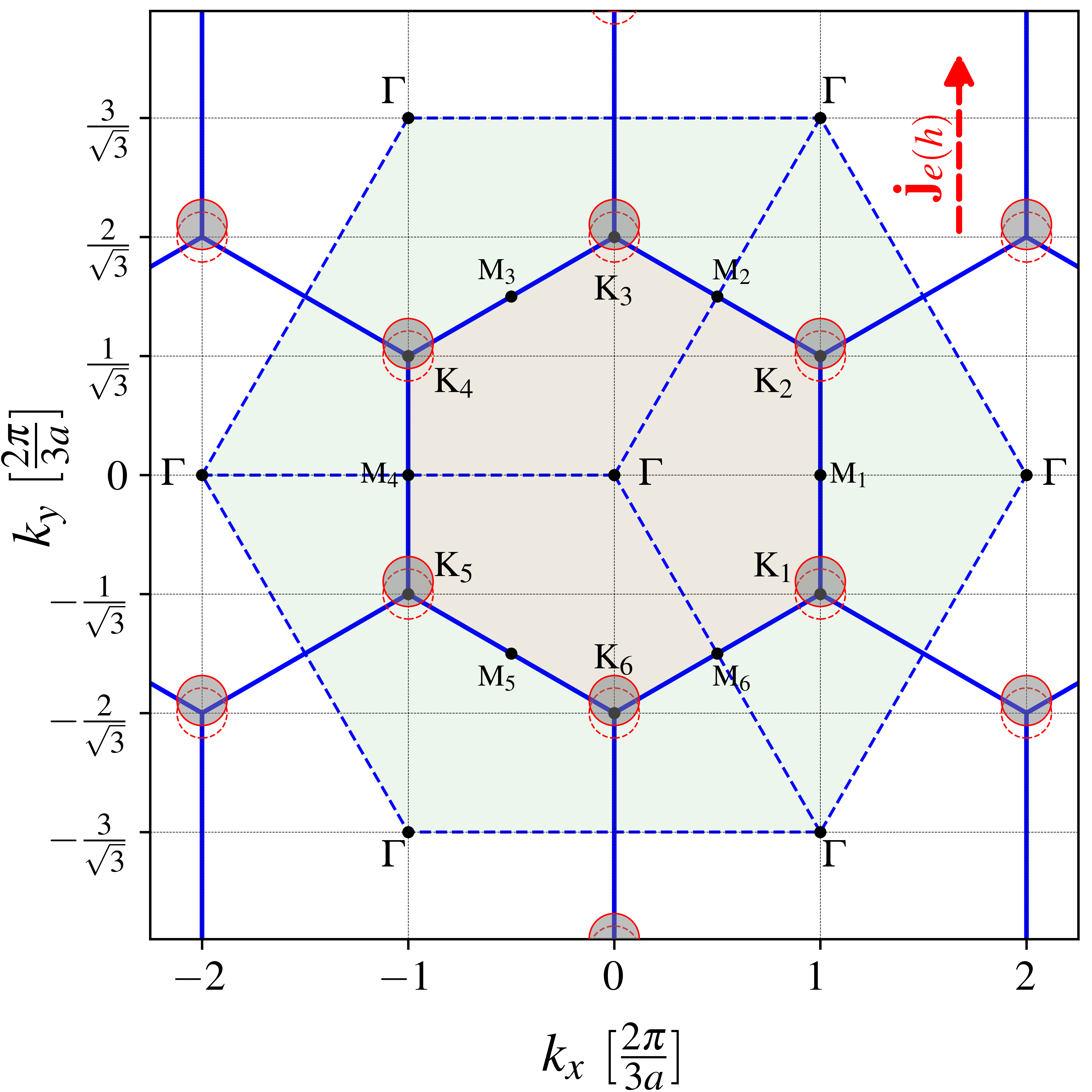}
		\caption{(Color online) The two-dimensional reciprocal space corresponding to the honeycomb lattice of carbon atoms in real space. The hexagon colored in light red shows the first Brillouin zone, and the three rhombi
colored in light green show the other choices of summation domain in computing the relevant quantities. The dashed circles represent the (un)occupied eigen-states in the (valence) conduction band of ($p$-)$n$-doped graphene. The gray-filled circles show their counterpart in the presence of DC current along the $\hat{\bm{y}}$ direction. In general, the shift of the circles, which is parallel to the DC current, does not have to be along $\hat{\bm{y}}$ and can be in any arbitrary direction. The horizontal and vertical axes respectively describe the $x$ and $y$ components of the crystal momentum, $\bm{k}$, corresponding to each point in $\bm{k}$-space in units of $2\pi / 3a$ with $a \cong 0.142\,\mathrm{nm}$ being the carbon-carbon bond length.}
		\label{FBZ}
	\end{center}
\end{figure}

As shown in Fig.~\ref{FBZ}, $\Gamma$ and $\mathrm{K}_{j}$ points ($j=1,\ldots,6$) represent the center and the six corners of the hexagonal first Brillouin zone (FBZ), respectively. The phonon modes of interest in this work include the in-plane \textbf{l}ongitudinal/\textbf{t}ransverse \textbf{o}ptical phonon modes at the $\Gamma$ point ($\Gamma\!\operatorname{-}\!\mathrm{LO}$/$\Gamma\!\operatorname{-}\!\mathrm{TO}$) and the in-plane \textbf{t}ransverse \textbf{o}ptical phonon modes at the $\mathrm{K}_{j}$ point ($\mathrm{K}_{j}\!\operatorname{-}\!\mathrm{TO}$) \footnote{We adopted the abbreviation method used in Ref.~\onlinecite{Malard_2009} to represent these phonon modes.}. Since the electron-phonon coupling (EPC) is large only for these modes \cite{Piscanec_2004}, the impact of the flowing $\pi$ electron gas, which can be regarded as a perturbation to EPC, is also expected to be non-negligible only for these modes. The out-of-plane optical (ZO) phonon modes are not expected to be affected by DC current due to their weak coupling to the Dirac fermions in graphene \cite{Samsonidze_2007,Basko_2008,Politano_2015}, and only in quasi-freestanding graphene epitaxially grown on Pt(111) the ZO branch has been reported to exhibit signatures of large EPC in the vicinity of the $\Gamma$ point \cite{Politano_2015}. The phonon modes which are the focus of this work are denoted in the $\left(\nu , \bm{q}\right)$ representation as follows,
\begin{eqnarray*}\label{MODE_NOTATION}
\Gamma\!\operatorname{-}\!\mathrm{LO} &:& \left(\mathrm{LO} , \bm{q} = \bm{0}\right),
\\[0.1ex]
\Gamma\!\operatorname{-}\!\mathrm{TO} &:& \left(\mathrm{TO} , \bm{q} = \bm{0}\right),
\\[0.1ex]
\mathrm{K}_{j}\!\operatorname{-}\!\mathrm{TO} &:& \left(\mathrm{TO} , \bm{q} = \mathbf{K}_{j}\right) \quad{;}j=1,\ldots,6.
\end{eqnarray*}
The interaction of a phonon mode, $\left(\nu , \bm{q}\right)$, with the $\pi$ electron gas leads to (i) the renormalization of its frequency, $\omega_{\nu, \bm{q}}$, and (ii) the emergence of a finite broadening, $\gamma_{\nu, \bm{q}}$. The creation and annihilation of electron-hole pairs in the $\pi$--$\pi^{*}$ bands arising from the scattering processes involving the $\left(\nu , \bm{q}\right)$ phonon mode leads to a phonon self-energy, $\Pi_{\nu,\bm{q}}$, and the electron-phonon interaction shall thus be discussed via $\Pi_{\nu,\bm{q}}$. The resulting frequency renormalization and uncertainty are expressed by \cite{Ando_2006,Araujo_2012},
\begin{align}
&\hbar\omega_{\nu, \bm{q}} = \hbar\omega_{\nu, \bm{q}}^{\mathrm{B}} + \mathrm{Re}\!\left[\Pi_{\nu,\bm{q}}\right]\label{PHONON_RENORMALIZATION_1}
\\[1.0ex]
&\hbar\gamma_{\nu, \bm{q}} = \hbar\gamma_{\nu, \bm{q}}^{\mathrm{B}} - \mathrm{Im}\!\left[\Pi_{\nu,\bm{q}}\right]\label{PHONON_RENORMALIZATION_2},
\end{align}
with $\omega_{\nu, \bm{q}}^{\mathrm{B}}$ and $\gamma_{\nu, \bm{q}}^{\mathrm{B}}$ being respectively the frequency and broadening of the $\left(\nu , \bm{q}\right)$ mode of undoped graphene wherein the eigen-states in the valence band are all occupied and the eigen-states in the conduction band are empty \cite{Ando_2006}. The quantity $\gamma_{\nu, \bm{q}}^{\mathrm{B}}$ denotes the \textit{residual} broadening of the $\left(\nu , \bm{q}\right)$ mode due to its involvement in scattering processes such as (i) phonon-phonon (anharmonic effects) \cite{Paulatto_2013}, (ii) phonon-impurity, and (iii) phonon-defect.

Introducing a small DC current to the $\pi$ electron gas can be modeled as a perturbation to the self-energy. Such current-induced perturbation, $\delta\Pi_{\nu,\bm{q}}$, can be obtained by subtracting the self-energy computed in the absence of DC current, $\Pi^{\0}_{\nu,\bm{q}}$, from its NE value \footnote{Here, we denote the current-induced perturbation to quantity $Q$ by $\delta Q$ which is defined as the value of the quantity computed in the presence of the DC current, $Q$, subtracted by the value of the quantity computed in the absence of the DC current, $Q^{\0}$, i.e., $\delta Q \equiv Q - Q^{\0}$.}, i.e.,
\begin{equation}\label{EPC_BREAKDOWN}
\delta\Pi_{\nu,\bm{q}} = \Pi_{\nu,\bm{q}} - \Pi^{\0}_{\nu,\bm{q}}.
\end{equation}
As a result, Eqs.~(\ref{PHONON_RENORMALIZATION_1}) and~(\ref{PHONON_RENORMALIZATION_2}) can be re-written into
\begin{align}
&\hbar\omega_{\nu, \bm{q}} = \overbrace{\hbar\omega^{\mathrm{B}}_{\nu, \bm{q}}  + \mathrm{Re}\!\left[\Pi^{\0}_{\nu,\bm{q}}\right]}^\text{$\hbar\omega_{\nu, \bm{q}}^{\0}$} + \mathrm{Re}\!\left[\delta\Pi_{\nu,\bm{q}}\right]\label{PHONON_RENORMALIZATION_3}
\\[1.0ex]
&\hbar\gamma_{\nu, \bm{q}} = \underbrace{\hbar\gamma^{\mathrm{B}}_{\nu, \bm{q}} - \mathrm{Im}\!\left[\Pi^{\0}_{\nu,\bm{q}}\right]}_\text{$\hbar\gamma_{\nu, \bm{q}}^{\0}$} - \mathrm{Im}\!\left[\delta\Pi_{\nu,\bm{q}}\right]\label{PHONON_RENORMALIZATION_4},
\end{align}
with $\omega_{\nu, \bm{q}}^{\0}$ and $\gamma_{\nu, \bm{q}}^{\0}$ being respectively the frequency and broadening of mode $\left(\nu , \bm{q}\right)$ in the absense of DC current.

Raman spectra of current-carrying graphene have been measured in Refs.~\onlinecite{Freitag_2009,Berciaud_2010,Yin_2014,Son_2017} where the variation in the position and/or bandwidth of the $\mathrm{G}$ and $\mathrm{G}^{\prime}$ peaks with respect to the drain-source voltage have been mainly attributed to Joule heating; however, no attempt has been made to isolate the impact of the electron flow on these Raman features. In principle, as the sample heats up due to Joule heating, such isolation could be achieved by continuously cooling down the sample to maintain a sample temperature independent of the drain-source voltage.

In this paper, we investigate the direct (non-thermal) contribution of DC electric current to the position and bandwidth of the Raman $\mathrm{G}$ peak at a given temperature. This contribution is solely due to the asymetric nature of the occupation of the eigen-states around Dirac cones by the $\pi$ electron gas in its current-carrying state. The in-plane optical phonon modes of graphene at the center and corners of its FBZ are known to be responsible for the $\mathrm{G}$ and $\mathrm{G}^{\prime}$ features, respectively \cite{Malard_2009}. Therefore, the impact of DC current on these Raman features can be quantified by computing the current-induced perturbation to the self-energy, $\delta\Pi_{\nu,\bm{q}}$, for the responsible phonon modes, i.e., the $\Gamma\!\operatorname{-}\!\mathrm{LO}$, $\Gamma\!\operatorname{-}\!\mathrm{TO}$ and $\mathrm{K}_{j}\!\operatorname{-}\!\mathrm{TO}$ modes ($j=1,\ldots,6$).

The generalized formalism to compute the phonon mode renormalization will be explicitly discussed in Section~\ref{sec:FOR}. In Sec.~\ref{sec:MOD_NEPEG}, a method will be introduced to approximately describe the occupation of the eigen-states within the valence and conduction bands by the flowing $\pi$ electron gas. In Sec.~\ref{sec:OPT_GAM}, the computation of the DC-current-induced frequency shift and broadening for the $\mathrm{LO}$ and $\mathrm{TO}$ modes at the $\Gamma$ point will be discussed along with the numerical results. The same analysis will be repeated for the $\mathrm{TO}$ modes at the $\mathrm{K}_{j}$ points ($j=1,\ldots,6$) in Sec.~\ref{sec:OPT_K}. The experimental manifestation of such current-induced modification to the these phonon modes will be discussed in Sec.~\ref{sec:RAMAN}. The findings of this paper will then be summarized in Sec.~\ref{sec:S&C} followed by concluding remarks.

%%%%%%%%%%%%%%%%%%%%%%%%%%%%%%%%%%%%%%%%%%%%%%%%%%%%%%%%%%%%%
%SECTION II: Self-energy formalism
%%%%%%%%%%%%%%%%%%%%%%%%%%%%%%%%%%%%%%%%%%%%%%%%%%%%%%%%%%%%%
\section{Self-energy formalism}\label{sec:FOR}

Within the second-order perturbation theory, the self-energy of the $\left(\nu , \bm{q}\right)$ phonon mode due to its interaction with the $\pi$ electron gas is given by the following integral over the FBZ ($i \equiv \sqrt{-1}$) \cite{Ando_2006,Kong_2008,Kong_2009,Araujo_2012},
\begin{equation}\label{SELF_ENERGY}
\begin{split}
&\Pi_{\nu,\bm{q}} = \frac{g_{\scriptscriptstyle{\mathrm{S}}}}{A_{\scriptscriptstyle{\mathrm{F}}\scriptscriptstyle{\mathrm{B}}\scriptscriptstyle{\mathrm{Z}}}} \int_{\scriptscriptstyle{\mathrm{F}}\scriptscriptstyle{\mathrm{B}}\scriptscriptstyle{\mathrm{Z}}} \! \mathrm{d}^{2}\bm{k} \sum_{s,s^{\prime}=\pm} \Big[ \mathrm{F}^{[\nu]}_{s,s^{\prime}}\! \left(\bm{k},\bm{q}\right)\qquad\qquad
\\
&\qquad\qquad\quad\;\; \frac{n_{\scriptscriptstyle{\mathrm{F}}}[E_{s^{\prime}}(\bm{k}\!+\!\bm{q}) , E_{\scriptscriptstyle{\mathrm{F}}}] - n_{\scriptscriptstyle{\mathrm{F}}}[E_{s}(\bm{k}) , E_{\scriptscriptstyle{\mathrm{F}}}]}{E_{s^{\prime}}(\bm{k}\!+\!\bm{q}) - E_{s}(\bm{k}) \!-\! \hbar\left[\omega_{\nu,\bm{q}}\!+\!i\gamma_{\nu,\bm{q}}\right]}\Big],
\end{split}
\end{equation}
with $g_{\scriptscriptstyle{\mathrm{S}}}\!=\!2$, $A_{\scriptscriptstyle{\mathrm{F}}\scriptscriptstyle{\mathrm{B}}\scriptscriptstyle{\mathrm{Z}}} = \frac{2}{3\sqrt{3}} \left[\frac{2\pi}{a}\right]^{2}$ and $n_{\scriptscriptstyle{\mathrm{F}}}[E , E_{\scriptscriptstyle{\mathrm{F}}}]$ denoting the spin degeneracy, the area of the FBZ and the Fermi-Dirac (FD) distribution function given by
\begin{equation}\label{Fermi_Dirac}
n_{\scriptscriptstyle{\mathrm{F}}}[E , E_{\scriptscriptstyle{\mathrm{F}}}]=\left[1+\exp{\!\left(\frac{E - E_{\scriptscriptstyle{\mathrm{F}}}}{k_{\scriptscriptstyle{\mathrm{B}}}T_{e}}\right)}\right]^{-1},
\end{equation}
where $T_{e}$ and $E_{\scriptscriptstyle{\mathrm{F}}}$ respectively denote the temperature and the Fermi energy of the $\pi$ electron gas, and $k_{\scriptscriptstyle{\mathrm{B}}}$ is the Boltzmann constant. The function $E_{s}(\bm{k})$ yields the energy eigen-value of the $\ket{\bm{k},s}$ eigen-state of the conduction ($s\!=\!+1$) or valence ($s\!=\!-1$) band. The tight-binding (TB) model yields $E_{s}(\bm{k})$ in terms of the hopping parameters corresponding to the nearest-neighbor (NN) and the next-nearest-neighbor (NNN) carbon atoms in graphene, respectively denoted by $t \cong 2.7 \, \mathrm{eV}$ and $t^{\prime} \cong -0.2 t$ \cite{NETO_2009},
\begin{equation}\label{TB_MODEL_ENERGY_EIGEN_VALUE}
E_{s}(\bm{k}) \cong s t \sqrt{3 + f\!\left(\bm{k}\right)} - t^{\prime} f\!\left(\bm{k}\right),
\end{equation}
where $f\!\left(\bm{k}\right)$ is given as follows \cite{NETO_2009},
\begin{equation}\label{THE_TB_FUNCTION}
f\!\left(\bm{k}\right) = 2 \cos{\!\Big[\sqrt{3} k_{y} a\Big]} + 4 \cos{\!\Big[\frac{\sqrt{3} k_{y} a}{2}\Big]}  \cos{\!\Big[\frac{3}{2} k_{x} a\Big]},
\end{equation}
with $a\cong 0.142\,\mathrm{nm}$ and $\bm{k} = k_{x} \hat{\bm{x}}+ k_{y} \hat{\bm{y}}$ respectively being the carbon-carbon bond length in graphene, and the crystal momentum vector measured with respect to the $\Gamma$ point. The expression given by Eq.~(\ref{THE_TB_FUNCTION}) is valid for the choice of unit vectors, $\hat{\bm{x}}$ and $\hat{\bm{y}}$, shown in Fig.~\ref{FBZ}.

In addition, $\mathrm{F}^{[\nu]}_{s,s^{\prime}}\! \left(\bm{k},\bm{q}\right) = |\langle \bm{k}+\bm{q} , s^{\prime}| \partial_{\nu , \bm{q}}V|\bm{k},s\rangle|^{2}$ denotes the electron-phonon scattering amplitude between the $\ket{\bm{k},s}$ and $\ket{\bm{k}+\bm{q} , s^{\prime}}$ eigen-states due to interaction with the $\left(\nu , \bm{q}\right)$ phonon mode, where $\partial_{\nu , \bm{q}}V$ is the derivative of the electronic Kohn-Sham potential with respect to the atomic displacement along the mode eigen-vector $\hat{\bm{\mathrm{e}}}_{\nu,\bm{q}}$ \cite{Piscanec_2004,Lazzeri_2006,Pisana,Hu_2022}. The coupling between electrons and phonons in graphene can be understood from the TB model perspective by noting that the carbon-carbon bond length in graphene is modulated by the phonon modes, and the scattering amplitude obtained from the TB model reflects the impact of the phonon-induced bond length modulation on the NN hopping parameter \cite{Ando_2006,Neto_2007,Stauber_2008,Sohier_2015}. In the following sections, the electron-phonon scattering amplitude corresponding to the $\left(\mathrm{LO} , \bm{q} = \bm{0}\right)$, $\left(\mathrm{TO} , \bm{q} = \bm{0}\right)$, and $\left(\mathrm{TO} , \bm{q} = \mathbf{K}_{j} \right)$ modes will be discussed in more detail and the self-energy integral given by Eq.~(\ref{SELF_ENERGY}) will be simplified accordingly.

Since $\omega_{\nu, \bm{q}}^{\mathrm{B}}$ is the mode frequency of neutral graphene, it already contains the contribution of the $\pi$ electron gas at ground state. Therefore, in applying Eq.~(\ref{PHONON_RENORMALIZATION_1}) to the case of neutral graphene at ground state, i.e., $\omega_{\nu, \bm{q}} = \omega_{\nu, \bm{q}}^{\mathrm{B}} $, the term $\mathrm{Re}\!\left[\Pi_{\nu,\bm{q}}\right]$ is expected to vanish. However, for undoped graphene ($E^{\0}_{\scriptscriptstyle{\mathrm{F}}} = 0$) at $T_{e} = 0\,\mathrm{K}$, the self-energy integral in Eq.~(\ref{SELF_ENERGY}) yields the following nonzero value,
\begin{equation}\label{THE_VIRTUAL_EXCITATIONS}
\Pi^{\scriptscriptstyle{\mathrm{V}}\scriptscriptstyle{\mathrm{E}}}_{\nu,\bm{q}} \cong -2\frac{g_{\scriptscriptstyle{\mathrm{S}}}}{A_{\scriptscriptstyle{\mathrm{F}}\scriptscriptstyle{\mathrm{B}}\scriptscriptstyle{\mathrm{Z}}}} \int_{\scriptscriptstyle{\mathrm{F}}\scriptscriptstyle{\mathrm{B}}\scriptscriptstyle{\mathrm{Z}}}{\frac{ \mathrm{F}^{[\nu]}_{+,-}\! \left(\bm{k},\bm{q}\right) \, \mathrm{d}^{2}\bm{k}}{E_{+}(\bm{k}\!+\!\bm{q}) - E_{-}(\bm{k})}},
\end{equation}
where VE stands for Virtual Excitations; a term used in Ref.~\onlinecite{Ando_2006} to refer to this contribution. Since the integration cutoff energy is much larger than $\hbar \omega_{\nu, \bm{q}}^{\mathrm{B}}$, for the majority of the integration domain $|E_{+}(\bm{k}+\bm{q}) - E_{-}(\bm{k})| \gg \hbar \omega_{\nu, \bm{q}}^{\mathrm{B}}$ is satisfied. For this reason, one can ignore $\omega_{\nu,\bm{q}}+i\gamma_{\nu,\bm{q}}$ in the denominator of the integrand in Eq.~(\ref{SELF_ENERGY}); hence the term virtual excitations. Therefore, to avoid double-counting in Eqs.~(\ref{PHONON_RENORMALIZATION_1}) and~(\ref{PHONON_RENORMALIZATION_2}), the self-energy given by Eq.~(\ref{SELF_ENERGY}) should be redefined according to $\Pi_{\nu,\bm{q}} \to \Pi_{\nu,\bm{q}} - \Pi^{\scriptscriptstyle{\mathrm{V}}\scriptscriptstyle{\mathrm{E}}}_{\nu,\bm{q}}$ \cite{Ando_2006,Kong_2008}.

As suggested by Eq.~(\ref{EPC_BREAKDOWN}), the current-induced perturbation to the self-energy can be obtained by subtracting the self-energy computed in the absence of DC current, $\Pi^{\0}_{\nu,\bm{q}}$, from its value computed in the presence of DC current, $\Pi_{\nu,\bm{q}}$. Clearly, since the contribution due to the virtual excitations does not depend on the presence of DC current, DC-current-induced perturbations do not contain any contribution due to the virtual excitations.

It is worth noting that the set of Eqs.~(\ref{PHONON_RENORMALIZATION_1}),~(\ref{PHONON_RENORMALIZATION_2}) and~(\ref{SELF_ENERGY}) could, in principle, be solved self-consistently. However, owing to the perturbative nature of self-energy, our results did not change considerably in the second iteration of calculations. Therefore, in calculating the self-energy integral, we start with $\omega_{\nu,\bm{q}} = \omega_{\nu, \bm{q}}^{\mathrm{B}}$ and $\gamma_{\nu,\bm{q}}=\gamma_{\nu, \bm{q}}^{\mathrm{B}}$ in Eq.~(\ref{SELF_ENERGY}), and the corrected values for $\omega_{\nu,\bm{q}}$ and $\gamma_{\nu,\bm{q}}$ obtained from Eqs.~(\ref{PHONON_RENORMALIZATION_1}) and~(\ref{PHONON_RENORMALIZATION_2}) will \textit{not} be inserted back into Eq.~(\ref{SELF_ENERGY}) for the second step of calculations.
%%%%%%%%%%%%%%%%%%%%%%%%%%%%%%%%%%%%%%%%%%%%%%%%%%%%%%%%%%%%%
%SECTION III: The non-equilibrium state of the flowing $\pi$ electron gas
%%%%%%%%%%%%%%%%%%%%%%%%%%%%%%%%%%%%%%%%%%%%%%%%%%%%%%%%%%%%%
\section{The non-equilibrium state of the flowing $\pi$ electron gas}\label{sec:MOD_NEPEG}
\subsection{Modeling the non-equilibrium occupation}\label{subsec:MTNEO}

\begin{figure}[t!]
	\begin{center}
		\includegraphics[width = \columnwidth]{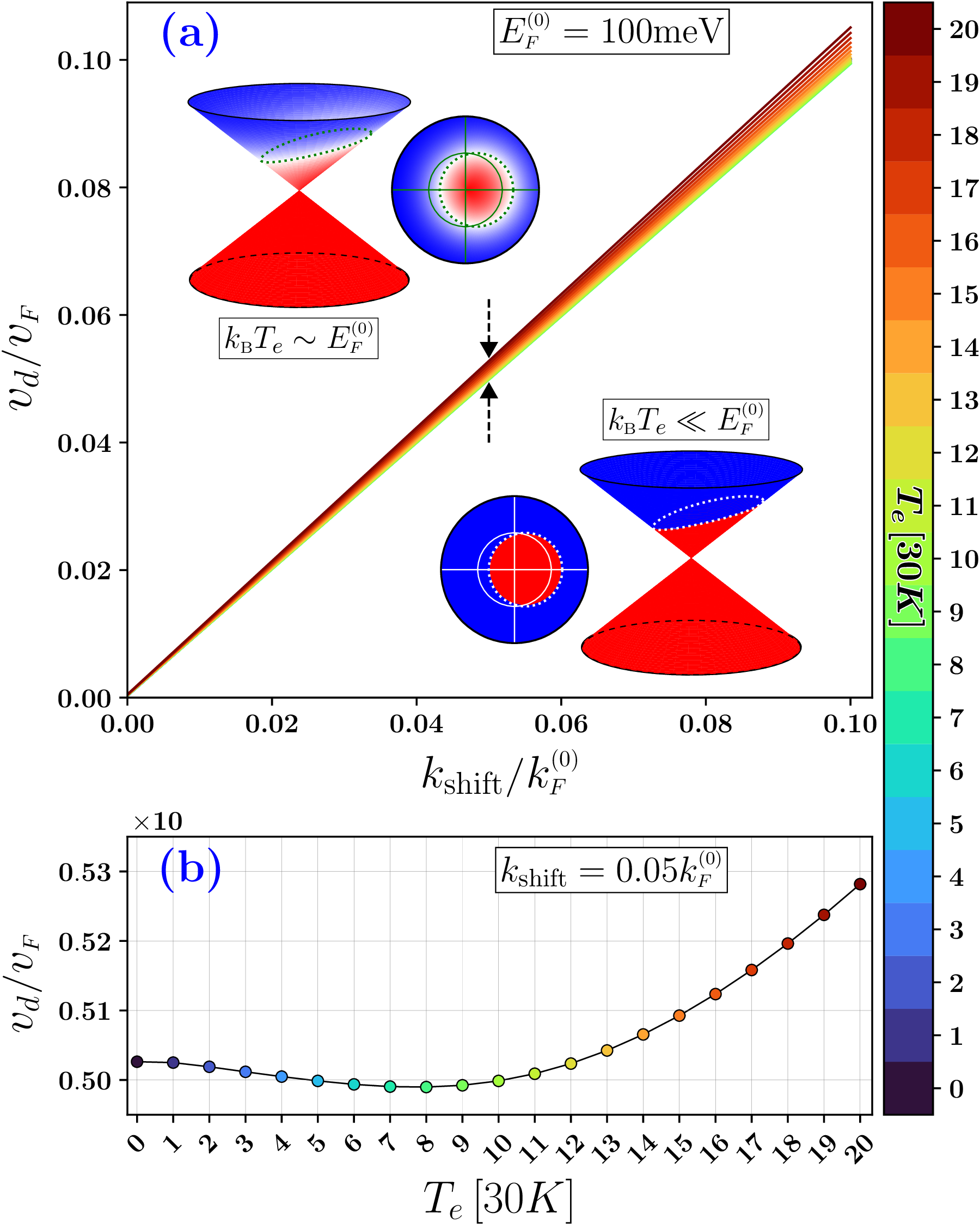}
		\caption{(Color online) The magnitude of the drift velocity, $v_{d}$, computed using Eq.~(\ref{Drift_Velocity}) for Fermi disk shift values up to $0.1 k_{\scriptscriptstyle{\mathrm{F}}}^{\0}$ for a graphene sample with a carrier density of $n_{s} \cong 10^{12}\mathrm{cm}^{-2}$. Panel \textbf{(a)} shows the $v_{d}$--$k_{\text{shift}}$ curves each generated assuming a fixed electron gas temperature, $T_{e}$, independent of the shift of the Fermi disk with $T_{e}$ ranging from $0.1\,\mathrm{K}$ up to $600\,\mathrm{K}$. The occupation of Dirac cones' electronic eigen-states has been illustrated at high and low temperatures wherein the (un)occupied eigen-states are shown with (blue) red, and the partially-occupied eigen-states are shown with white. Panel \textbf{(b)} displays the dependence of the drift velocity on $T_{e}$ at $k_{\text{shift}}= 0.05 k_{\scriptscriptstyle{\mathrm{F}}}^{\0}$, where the thermal spread of the drift velocity values is marked with two arrows on panel \textbf{(a)}.}
		\label{VD}
	\end{center}
\end{figure}

Similar to the approach taken in Refs.~\onlinecite{Sabstauber_2015,Sabstauber_2018}, in the absence of DC electric current, the FD distribution function in the phonon self-energy integral given by Eq.~(\ref{SELF_ENERGY}) should be used with the equilibrium-state Fermi energy, i.e., $E_{\scriptscriptstyle{\mathrm{F}}}^{\0} \!=\! \pm \hbar v_{\scriptscriptstyle{\mathrm{F}}} \sqrt{\pi n_{s}}$, with $n_{s}$ being the density of electrons injected into ($E^{\0}_{\scriptscriptstyle{\mathrm{F}}}\!>\!0$) or pulled out of ($E^{\0}_{\scriptscriptstyle{\mathrm{F}}}\!<\!0$) the graphene sample, and $\hbar v_{\scriptscriptstyle{\mathrm{F}}}$ is the slope of the Dirac cones. Applying drain-source voltage along the graphene channel drives the $\pi$ electron gas out of equilibrium, and the resulting NE occupation of the eigen-states in the valence and conduction bands can be calculated using the Boltzmann transport equation (BTE) \cite{Levinson,MAHAN_BTE,Vasko}. A simpler alternative approach is to employ the phenomenological shifted Fermi disk (SFD) model \cite{datta_1995} which estimates the NE electronic occupation of current-carrying graphene with the FD distribution function when used with an angle-dependent Fermi energy. This model consistently describes the occupation of the electronic eigen-states in the vicinity of the FBZ corners. To elaborate, we first define $\theta_{\bm{k}}$ to be the angle between the crystal momentum vector relative to the FBZ corner at $\mathbf{K}_{j}$ ($j = 1,\ldots,6$) and $\hat{\bm{x}}$, i.e., under the $\bm{k} \to \bm{k} - \mathbf{K}_{j}$ redefinition,
\begin{equation}\label{THETA_K}
\bm{k} = k \left[ \hat{\bm{x}} \cos{\theta_{\bm{k}}} + \hat{\bm{y}} \sin{\theta_{\bm{k}}} \right].
\end{equation}
The SFD model simulates the NE occupation with a shift of the Fermi disk, $\bm{k}_{\text{shift}}$, with respect to the corners of the hexagonal FBZ. This shift occurs parallel to the electron (hole) drift velocity, $\bm{v}_{d}\!=\!v_{d} \left[ \hat{\bm{x}} \cos{\theta_{d}} + \hat{\bm{y}} \sin{\theta_{d}} \right]$, for positive (negative) values of Fermi energy, $E_{\scriptscriptstyle{\mathrm{F}}}^{\0}$, i.e.,
\begin{equation}\label{K_SHIFT}
\bm{k}_{\text{shift}} = \eta k_{\scriptscriptstyle{\mathrm{F}}}^{\0} \left[ \hat{\bm{x}} \cos{\theta_{d}} + \hat{\bm{y}} \sin{\theta_{d}} \right],
\end{equation}
where $\eta \equiv k_{\text{shift}} / k_{\scriptscriptstyle{\mathrm{F}}}^{\0} \! \leq \! 1$ is the shift of the Fermi disk in units of the Fermi wavevector, $k_{\scriptscriptstyle{\mathrm{F}}}^{\0} \! \equiv \! |E_{\scriptscriptstyle{\mathrm{F}}}^{\0}| / [\hbar v_{\scriptscriptstyle{\mathrm{F}}}]$. Such a shift leads to a $\theta_{\bm{k}}$-dependent NE Fermi energy \cite{Sabstauber_2015,Sabstauber_2018},
\begin{equation}\label{Nonequilibrium_EF}
\frac{E_{\scriptscriptstyle{\mathrm{F}}}(\theta_{\bm{k}},\theta_{d})}{E_{\scriptscriptstyle{\mathrm{F}}}^{\0}}=\eta \cos{[\theta_{\bm{k}}\!-\!\theta_{d}]} + \sqrt{1-\eta^{2} \sin^{2}{\![\theta_{\bm{k}}\!-\!\theta_{d}]}}.
\end{equation}
The drift velocity of electrons/holes in graphene, $\bm{v}_{d}$, can then be calculated from \cite{Sabstauber_2018},
\begin{equation}\label{Drift_Velocity}
\bm{v}_{d} \!=\! \frac{v_{\scriptscriptstyle{\mathrm{F}}}}{\pi^{2} n_{s}} \int_{0}^{2\pi}{\!\!\!\int_{0}^{k_{c}}{\!\bm{k} \, n_{\scriptscriptstyle{\mathrm{F}}}\!\left[\hbar v_{\scriptscriptstyle{\mathrm{F}}} k , E_{\scriptscriptstyle{\mathrm{F}}}(\theta_{\bm{k}},\theta_{d})\right] \mathrm{d}k \, \mathrm{d}\theta_{\bm{k}}}},
\end{equation}
with $\mathrm{sgn}\!\left[x\right]$, $v_{\scriptscriptstyle{\mathrm{F}}} = \frac{3at}{2\hbar}$ and $k_{c} = 2k_{\scriptscriptstyle{\mathrm{F}}}^{\0} + \frac{5}{3at}k_{\scriptscriptstyle{\mathrm{B}}}T_{e}$ respectively denoting the signum function, the Fermi velocity, and the cutoff for the radial $k$-integration. The drift velocity is converted to the surface current density, $\bm{j}_{s}$, by
\begin{equation}\label{CURRENT_DENSITY}
\bm{j}_{s} = \mathrm{sgn}\!\left[E_{\scriptscriptstyle{\mathrm{F}}}^{\0}\right] j_{\scriptscriptstyle{\mathrm{F}}} \frac{\bm{v}_{d}}{v_{\scriptscriptstyle{\mathrm{F}}}},
\end{equation}
where $j_{\scriptscriptstyle{\mathrm{F}}} \equiv e n_{s} v_{\scriptscriptstyle{\mathrm{F}}}$ and $e$ denotes the charge of an electron. In the $\eta \sim 1$ limit, where the shift of the Fermi disk is comparable to its radius, the $\theta_{\bm{k}}$-dependent Fermi energy that is formulated by Eq.~(\ref{Drift_Velocity}) fails to give a consistent description of the electronic occupation in both conduction and valence bands at finite temperatures. For this reason, we restrict our use of Eq.~(\ref{Drift_Velocity}) to the case where the shift of Fermi disk is small, i.e., $\eta^{2} \! \ll \! 1$. In this limit, Eq.(\ref{Nonequilibrium_EF}) can be simplified into
\begin{equation}\label{Nonequilibrium_kF_SIMPLIFIED}
k_{\scriptscriptstyle{\mathrm{F}}}(\theta_{\bm{k}},\theta_{d}) = k_{\scriptscriptstyle{\mathrm{F}}}^{\0} \left[1 + \eta \cos{\left(\theta_{\bm{k}}\!-\!\theta_{d}\right)} \right] + \mathcal{O}\!\left[\eta^{2}\right].
\end{equation}
Figure~\ref{VD} shows the drift velocity computed with Eqs.~(\ref{Nonequilibrium_EF}) and~(\ref{Drift_Velocity}) for modest values of $k_{\text{shift}}$, and we observe that
\noindent the computed drift velocity still exhibits a linear behaviour versus $k_{\text{shift}}$ at temperatures as high as $600\,\mathrm{K}$. This justifies that the terms proportional to $\eta^{2}$ or higher-order terms in Eq.~(\ref{Nonequilibrium_kF_SIMPLIFIED}) can safely be discarded.
\subsection{The experimental relevance of the SFD model}\label{subsec:TEROTM}

Here, we present a series of simplified arguments to relate the experimental parameters such as drain-source voltage, gate voltage, and some of the geometrical parameters of a graphene Field-Effect Transistor (G-FET) to the parameters introduced in the SFD model such as $k_{\text{shift}}$ and $E^{\0}_{\scriptscriptstyle{\mathrm{F}}}$. The shift of the Fermi disk can be obtained from the channel length, $l_{c}$, the drain-source voltage, $V_{\mathrm{ds}}$, and momentum relaxation time $\tau_{m}$ \cite{datta_1995},
\begin{equation}\label{kshift}
\eta = \frac{k_{\text{shift}}}{k_{\scriptscriptstyle{\mathrm{F}}}^{\0}} \cong \frac{e V_{\mathrm{ds}}}{\hbar l_{c}} \frac{\tau_{m}}{\sqrt{\pi n_{s}}}.
\end{equation}
For a momentum relaxation time of $\tau_{m} = 60\,\mathrm{fs}$ \cite{Condori_2015}, a carrier density of $n_{s} = 10^{12}\mathrm{cm}^{-2}$, a drain-source voltage of $V_{\mathrm{ds}} = 0.1\mathrm{V}$ and a channel length of $l_{c} = 5\mathrm{\mu m}$, Eq.~(\ref{kshift}) yields $\eta \cong 0.01$. We can obtain the channel mobility $\mu_{c}$ which corresponds to the momentum relaxation time of $\tau_{m} = 60\,\mathrm{fs}$. As shown in Fig.~\ref{VD}, in the low-current regime the drift velocity can be safely described by $v_{d} \cong \eta v_{\scriptscriptstyle{\mathrm{F}}}$. Therefore, the channel mobility reads
\begin{equation}\label{channel_mobility}
\mu_{c} \equiv \dfrac{v_{d}}{V_{\mathrm{ds}}} l_{c} \cong \frac{e\tau_{m}}{\hbar \sqrt{\pi n_{s}}} \times \frac{3at}{2\hbar},
\end{equation}
which yields a channel mobility of $\mu_{c} \cong 4500 \,\frac{\mathrm{cm}^{2}}{\mathrm{V}\cdot \mathrm{s}}$. On the other hand, the normalized drift velocity values shown in Fig.~\ref{VD} can be converted to a value for surface current density when multiplied by $j_{\scriptscriptstyle{\mathrm{F}}} = e n_{s} v_{\scriptscriptstyle{\mathrm{F}}} \cong 1.4 \, \mathrm{mA}/\mathrm{\mu m}$. Multiplying the resulting surface current density by the channel width $w_{c}$ yields the total current that flows through the channel. For $\eta \cong 0.01$, the curves presented in Fig.~\ref{VD} yield $v_{d}/v_{\scriptscriptstyle{\mathrm{F}}} \cong 0.01$, which corresponds to a channel current of $I_{c} \cong 70\,\mu\mathrm{A}$ for a graphene ribbon of width $w_{c} = 5\mu\mathrm{m}$ \footnote{Comparing the numerical values discussed here as an example, i.e., $V_{\mathrm{ds}} = 0.1\mathrm{V} / [5\mu\mathrm{m}] =  0.2\mathrm{KV}/\mathrm{cm}$ and $|\bm{j}_{s}| \cong \eta j_{\scriptscriptstyle{\mathrm{F}}} = 14 \mu\mathrm{A}/\mu\mathrm{m}$, with the I-V characteristics obtained from the numerical solution of the BTE in Ref.~\onlinecite{Chauhan_2009} indicates that the SFD model is in agreement with the BTE in the linear regime of the I-V curves wherein the drain-source voltage is relatively small. Moreover, the agreement between the SFD model and the experimental I-V characteristics in the low-bias regime can be seen in Refs.~\onlinecite{Meric_2008,Barreiro_2009}.}. Also, we can find an estimate for the gate voltage, $V_{g}$, needed to achieve a certain carrier density,
\begin{equation}\label{GATE_MODEL}
\frac{V_{g}}{t_{g}} \cong \frac{e n_{s}}{\epsilon_{g} \varepsilon_{0}},
\end{equation}
with $\epsilon_{g}$, $ \varepsilon_{0}$ and $t_{g}$ respectively being the gate dielectric constant, the permittivity of vacuum and the gate thickness. The expression given by Eq.~(\ref{GATE_MODEL}) is obtained based on a simplified approach which models the gate and the channel as two plates of a capacitor. Assuming the thickness and the dielectric constant of the gate to be $t_{g} = 300\,\mathrm{nm}$ and $\epsilon_{g} = 3$ \cite{Freitag_2009,Bresnehan_2012,Lee_2013,Meric_2013,Mohrmann_2014}, the expression given by Eq.~(\ref{GATE_MODEL}) suggests that a gate voltage of $V_{g} \cong 18\,\mathrm{V}$ is required to achieve a carrier density of $n_{s} = 10^{12}\, \mathrm{cm}^{-2}$. For a comprehensive study of G-FET devices, see Ref.~\onlinecite{Meric_dissertation}.

In the following sections the results will be reported in terms of the shift parameter $\eta$ and the equilibrium-state Fermi energy, $E_{\scriptscriptstyle{\mathrm{F}}}^{\0}$, rather than channel current density, the drain-source voltage and/or the gate voltage.

%%%%%%%%%%%%%%%%%%%%%%%%%%%%%%%%%%%%%%%%%%%%%%%%%%%%%%%%%%%%%
%SECTION IV: Current-induced perturbation to the in-plane optical phonon modes at the FBZ center
%%%%%%%%%%%%%%%%%%%%%%%%%%%%%%%%%%%%%%%%%%%%%%%%%%%%%%%%%%%%%
\section{Current-induced perturbation to the in-plane optical phonon modes at the FBZ center}\label{sec:OPT_GAM}
\subsection{Scattering amplitude formalism}\label{subsec:FOR_IV}
The electron-phonon scattering amplitude for the $\left(\nu , \bm{q}\right)$ phonon modes in the vicinity of the center of FBZ, i.e., $\left|\bm{q}\right| = q \ll a^{-1}$, is given as \cite{Lazzeri_2005,Ando_2006,Butscher_2007,Kong_2008,Kong_2009,Sasaki_2012,Tomadin_2013,Park_2014,Sohier_2014,Sohier_2015},
\begin{equation}\label{ELPH_SA_GAMMA}
\mathrm{F}^{[\nu]}_{s,s^{\prime}}\! \left(\bm{k},\bm{q}\right) = \kappa^{2}_{\Gamma} \left[\frac{1 - l_{\nu} \, ss^{\prime} \cos{\left(2\phi_{\bm{k} , \bm{q}}\right)}}{2}\right],
\end{equation}
where $\nu = \left\{\mathrm{LO} , \mathrm{TO}\right\}$, and $l_{\nu}$ is a mode index which takes the values of $+1$ and $-1$ for $\nu = \mathrm{LO}$ and $\nu = \mathrm{TO}$, respectively. Moreover, $\phi_{\bm{k} , \bm{q}}$ is defined as
\begin{equation}\label{PHI_DEFINITION}
\phi_{\bm{k} , \bm{q}} = \frac{\theta_{\bm{k} + \bm{q}} + \theta_{\bm{k}}}{2} - \theta_{\bm{q}},
\end{equation}
where $\theta_{\bm{k}} \!\equiv\! \angle{\left(\bm{k},\hat{\bm{x}}\right)}$, $\theta_{\bm{q}} \!\equiv\! \angle{\left( \bm{q},\hat{\bm{x}} \right)}$ and $\theta_{\bm{k} + \bm{q}} \!\equiv\! \angle{\left( \bm{k} \!+\! \bm{q},\hat{\bm{x}} \right)}$ \footnote{$\angle{\left(\bm{u} , \bm{v} \right)}$ denotes the angle between the vectors $\bm{u}$ and $\bm{v}$.}. The coupling parameter $\kappa_{\Gamma}$ for the $\nu = \left\{\mathrm{LO} ,  \mathrm{TO} \right\}$ modes is given by the TB model as \cite{Ando_2006,Kong_2009,Park_2014,Sohier_2014,Sohier_2015},
\begin{equation}\label{COUPLING_PARAMETER}
\kappa_{\Gamma} = \frac{3t\beta}{\sqrt{2} a} \sqrt{\frac{\hbar}{2M \omega^{\0}_{\Gamma}}} \cong 0.24 \, \mathrm{eV},
\end{equation}
where $M\cong 11.178 \, \mathrm{GeV}\!/c^{2}$ is the mass of the $^{12}_{\;\,6}\mathrm{C}$ carbon isotope and $\hbar\omega^{\0}_{\Gamma} \!=\! \hbar\omega^{\0}_{\mathrm{LO},\bm{0}} \!=\! \hbar\omega^{\0}_{\mathrm{TO},\bm{0}} \!\cong\! 196 \, \mathrm{meV}$ \cite{Piscanec_2004,Lazzeri_2005,Ando_2006,Piscanec_2007}, or equivalently $1581\,\mathrm{cm}^{-1}$, is the frequency of the degenerate $\left(\mathrm{LO} , \bm{q} = \bm{0}\right)$ and $\left(\mathrm{TO} , \bm{q} = \bm{0} \right)$ modes in the absence of DC current. The measured values for the frequency of these two modes in graphite range from $1565$ to $1583\,\mathrm{cm}^{-1}$ \cite{Maultzsch_2004}. Nevertheless, we take $1581\,\mathrm{cm}^{-1}$ as the nominal mode frequency. Finally, $\beta$ is a parameter which reflects the change in the NN hopping parameter due to the change in the bond length \cite{Ando_2006,Wendler_2015},
\begin{equation}\label{BETA}
\beta \equiv - \frac{d\left[\ln{t}\right]}{d\left[\ln{a}\right]}.
\end{equation}
Following Ref.~\onlinecite{Ando_2006}, all the numerical results in this paper are calculated for $\beta = 2$. One method to obtain $\beta$ is measuring the step-like decrease of the line-width of the Raman $\mathrm{G}$ peak upon increasing the carrier density with no DC current present \cite{Yan_2007,YAN200739}. Depending on the experimental dataset at hand, one could obtain a slightly different value for $\beta$, and in that case, all the numerical values for the phonon mode renormalizations and current-induced perturbations reported in our work can be corrected upon multiplication by $\beta^{2}/4$.

In the absence of DC current, graphene can be regarded as isotropic for very large phonon wavelengths, i.e., $\bm{q}\!=\!\bm{0}$, and therefore any set of mutually-orthogonal in-plane unit vectors can be selected as the eigen-vectors of the dynamical matrix at $\bm{q}\!=\!\bm{0}$. However, the presence of DC current breaks this isotropy, and the eigen-vectors of the perturbed dynamical tensor at $\bm{q} \!=\! \bm{0}$ are expected to be constructed from the unit vector defined by the introduced preferential direction. As a result, the in-plane atomic vibrations corresponding to the $\left(\mathrm{LO} , \bm{q} \!=\! \bm{0}\right)$ and $\left(\mathrm{TO} , \bm{q} \!=\! \bm{0} \right)$ modes should be parallel and perpendicular to the surface current density, respectively (For a detailed discussion, see Appendix~\ref{DOMEVITPODC}). As a consequence, $\theta_{\bm{q}}$ in the equilibrium-state scattering amplitude given by Eq.~(\ref{ELPH_SA_GAMMA}) should be replaced with $\Theta = \theta_{d} \pm \pi$ at $\bm{q} \!=\! \bm{0}$, and the expression for the scattering amplitude of the $\nu = \left\{\mathrm{LO}, \mathrm{TO}\right\}$ modes of current-carrying graphene reads,
\begin{equation}\label{ELPH_SA_GAMMA_REDEFINED}
\mathrm{\tilde{F}}^{[\nu]}_{s,s^{\prime}}\! \left(\bm{k},\bm{0}\right) = \frac{\kappa^{2}_{\Gamma}}{2} \Big\{1 - l_{\nu} \, ss^{\prime} \cos{\left(2\left[\theta_{\bm{k}} - \theta_{d}\right]\right)}\Big\}.
\end{equation}
Computing the self-energy integral with the preceding scattering amplitude leads to corrections for the $\left(\mathrm{LO} , \bm{q} \!=\! \bm{0}\right)$ and $\left(\mathrm{TO} , \bm{q} \!=\! \bm{0}\right)$ modes which are independent of the direction of the DC current. This can be verified by performing the angular integration over the $\theta$ variable obtained by implementing the change-of-variable
\begin{figure}[t!]
	\begin{center}
		\includegraphics[width = \columnwidth]{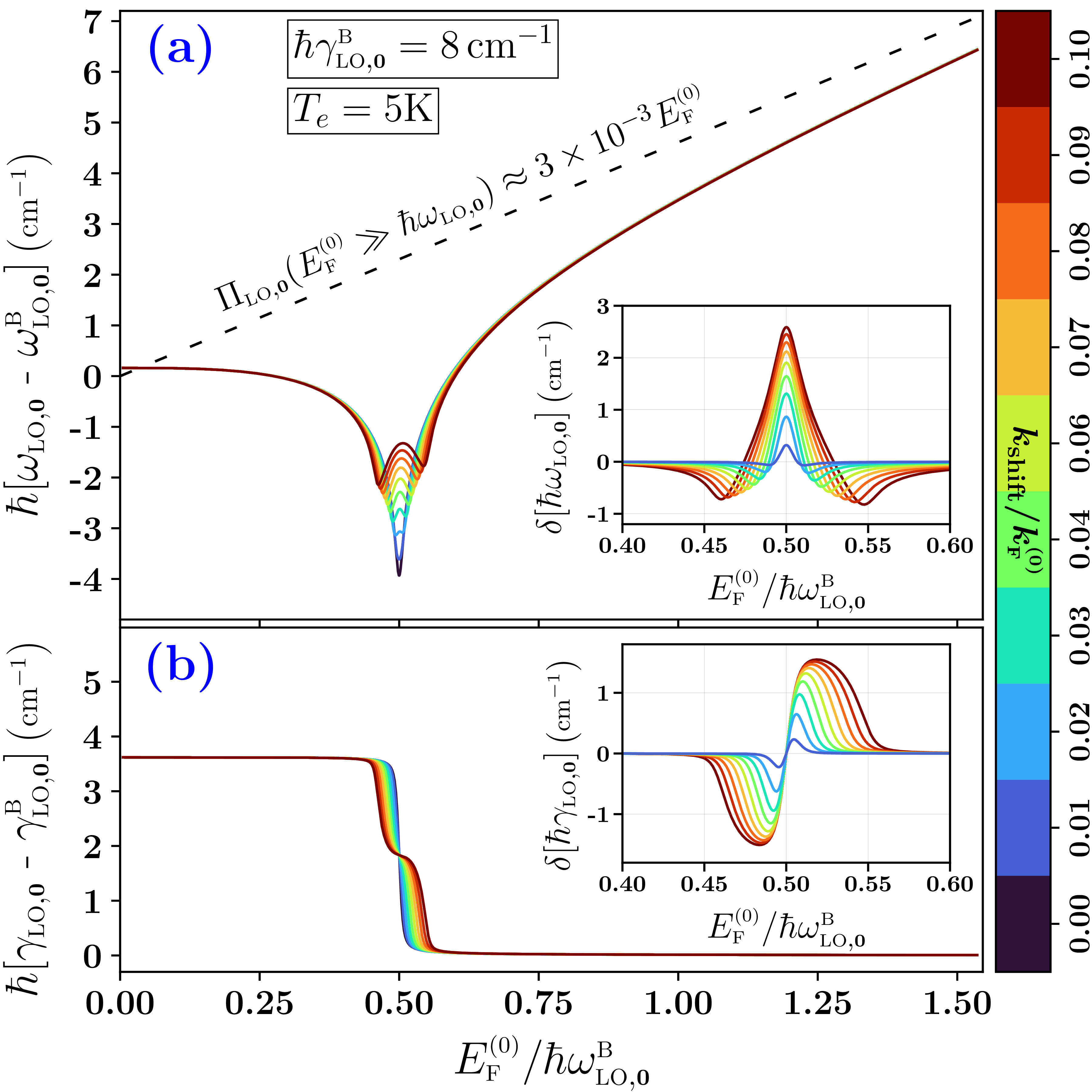}
		\caption{(Color online) Self-energy corrections obtained for the $\Gamma\!\operatorname{-}\!\mathrm{LO}$ mode by combining Eqs.~(\ref{SELF_ENERGY})--(\ref{THE_TB_FUNCTION}),~(\ref{Nonequilibrium_EF}),~(\ref{COUPLING_PARAMETER}), (\ref{BETA}),~(\ref{ELPH_SA_GAMMA_REDEFINED}) and~(\ref{VE_BG}) for Fermi disk shift values up to $0.1 k_{\scriptscriptstyle{\mathrm{F}}}^{\0}$ at $T_{e} \!=\! 5\mathrm{K}$. Panels \textbf{(a)} and \textbf{(b)} show respectively the frequency and broadening of the $\Gamma\!\operatorname{-}\!\mathrm{LO}$ mode of current-carrying graphene versus the equilibrium-state Fermi energy normalized by the frequency of the $\Gamma\!\operatorname{-}\!\mathrm{LO}$ mode of undoped graphene. The inset in each panel shows the current-induced perturbation to the quantity shown in its respective panel versus the normalized $E_{\scriptscriptstyle{\mathrm{F}}}^{\0}$. The Fermi energy is varied from $0$ to $300\,\mathrm{meV}$, and the current-induced perturbations exhibit a resonant-like behavior at a sample carrier density corresponding to $E_{\scriptscriptstyle{\mathrm{F}}}^{\0} = \hbar \omega^{\0}_{\Gamma}/2 \cong 98\,\mathrm{meV}$. In the presence of DC current, the $\Gamma\!\operatorname{-}\!\mathrm{LO}$ mode is defined as the mode with an in-plane atomic displacement \textit{parallel} to DC current flow.}
		\label{LO_VD}
	\end{center}
\end{figure}
\noindent given by $[\theta_{\bm{k}} - \theta_{d}] \to \theta$ in Eqs.~(\ref{Nonequilibrium_EF}), and~(\ref{ELPH_SA_GAMMA_REDEFINED}). This agrees with what one would expect intuitively because the intra-valley transitions caused by the $\left(\mathrm{LO} , \bm{q} \!=\! \bm{0}\right)$ and $\left(\mathrm{TO} , \bm{q} \!=\! \bm{0}\right)$ modes are strictly vertical and should not be affected by the direction of DC current.

A similar approach of utilizing the correct ``longitudinal'' and ``transverse'' polarizations has been taken in Refs.~\onlinecite{Mohiuddin_2009,Huang_2009,Bissett_2014} to interpret the $\mathrm{G}$ peak features of the Raman spectra for graphene samples under uniaxial strain with the in-plane atomic displacement of the $\Gamma\!\operatorname{-}\!\mathrm{LO}$ and $\Gamma\!\operatorname{-}\!\mathrm{TO}$ modes being parallel and perpendicular to the strain axis, respectively. Neglecting the correct polarization of the $\Gamma\!\operatorname{-}\!\mathrm{LO}$ and $\Gamma\!\operatorname{-}\!\mathrm{TO}$ modes relative to the preferred axis in graphene plane leads to a frequency surface that is not single-valued at $\bm{q} = \bm{0}$. For example, unlike Refs.~\onlinecite{Mohiuddin_2009,Huang_2009,Bissett_2014}, Ref.~\onlinecite{Assili_2014} did not utilize the correct basis set for the eigen-vectors of the $\Gamma\!\operatorname{-}\!\mathrm{LO}$ and $\Gamma\!\operatorname{-}\!\mathrm{TO}$ modes in the case of graphene under uniaxial strain, which led to the dependence of the self-energy corrections on the angle of the
\begin{figure}[t!]
	\begin{center}
		\includegraphics[width = \columnwidth]{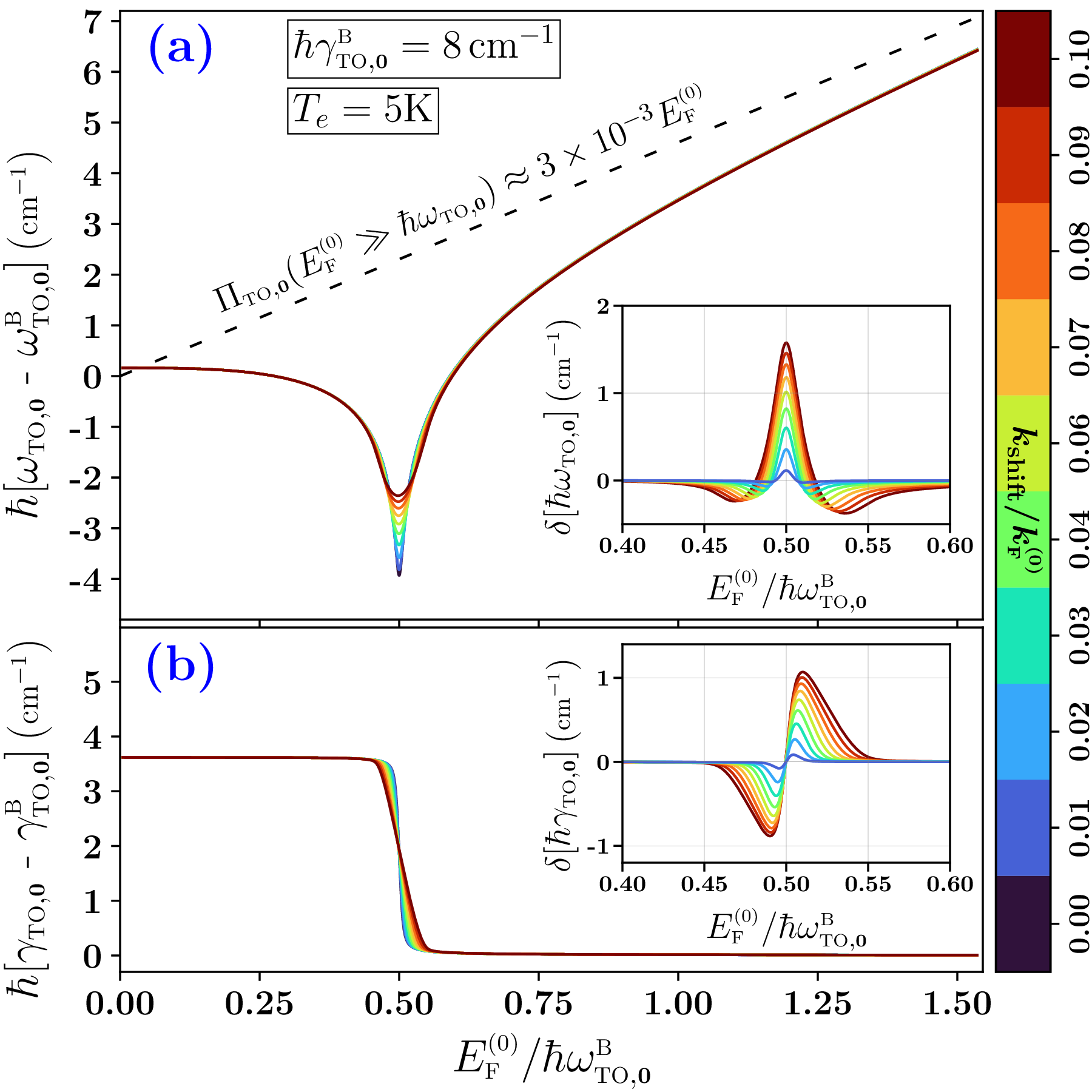}
		\caption{(Color online) Self-energy corrections obtained for the $\Gamma\!\operatorname{-}\!\mathrm{TO}$ mode by combining Eqs.~(\ref{SELF_ENERGY})--(\ref{THE_TB_FUNCTION}),~(\ref{Nonequilibrium_EF}),~(\ref{COUPLING_PARAMETER}), (\ref{BETA}),~(\ref{ELPH_SA_GAMMA_REDEFINED}) and~(\ref{VE_BG}) for Fermi disk shift values up to $0.1 k_{\scriptscriptstyle{\mathrm{F}}}^{\0}$ at $T_{e} \!=\! 5\mathrm{K}$. Panels \textbf{(a)} and \textbf{(b)} show respectively the frequency and broadening of the $\Gamma\!\operatorname{-}\!\mathrm{TO}$ mode of current-carrying graphene versus the equilibrium-state Fermi energy normalized by the frequency of the $\Gamma\!\operatorname{-}\!\mathrm{TO}$ mode of undoped graphene. The inset in each panel shows the current-induced perturbation to the quantity shown in its respective panel versus the normalized $E_{\scriptscriptstyle{\mathrm{F}}}^{\0}$. The Fermi energy is varied from $0$ to $300\,\mathrm{meV}$, and the current-induced perturbations exhibit a resonant-like behavior at a sample carrier density corresponding to $E_{\scriptscriptstyle{\mathrm{F}}}^{\0} = \hbar \omega^{\0}_{\Gamma}/2 \cong 98\,\mathrm{meV}$. In the presence of DC current, the $\Gamma\!\operatorname{-}\!\mathrm{TO}$ mode is defined as the mode with an in-plane atomic displacement \textit{perpendicular} to DC current flow.}
		\label{TO_VD}
	\end{center}
\end{figure}
mode momentum, $\bm{q}$, even when $q \!=\! 0$.

Figures~\ref{LO_VD} and~\ref{TO_VD} show the frequency and broadening of respectively the $\Gamma\!\operatorname{-}\!\mathrm{LO}$ and $\Gamma\!\operatorname{-}\!\mathrm{TO}$ modes in the presence of DC current for several values of DC current, with the current-induced perturbations being shown in the insets. The mode frequencies shown in Figs.~\ref{LO_VD}-\textbf{(a)} and~\ref{TO_VD}-\textbf{(a)} are obtained by subtracting the contribution of virtual excitations from the computed self-energy corrections. Within the Dirac cone approximation, the contribution due to the virtual excitations can be obtained from Eq.~(\ref{THE_VIRTUAL_EXCITATIONS}), and the resulting expression reads
\begin{equation}\label{VE_BG}
\Pi^{\scriptscriptstyle{\mathrm{V}}\scriptscriptstyle{\mathrm{E}}}_{\mathrm{LO},\bm{0}} = \Pi^{\scriptscriptstyle{\mathrm{V}}\scriptscriptstyle{\mathrm{E}}}_{\mathrm{TO}, \bm{0}} = - \left[\frac{g_{\scriptscriptstyle{\mathrm{S}}}g_{\scriptscriptstyle{\mathrm{V}}}}{4}\right] \frac{\sqrt{3}}{\pi t} \kappa^{2}_{\Gamma} \left[k_{c}a\right],
\end{equation}
with $g_{\scriptscriptstyle{\mathrm{V}}} \!=\! 2$ and $k_{c}$ being respectively the valley degeneracy and the cutoff used in evaluating the self-energy integral for the $\Gamma\!\operatorname{-}\!\mathrm{LO}$ and $\Gamma\!\operatorname{-}\!\mathrm{TO}$ modes. In our computations, we utilized $k_{c}= [10 k_{\scriptscriptstyle{\mathrm{B}}}T_{e} + 8 \hbar\omega^{\0}_{\Gamma}] / [\hbar v_{\scriptscriptstyle{\mathrm{F}}}]$.

The frequency and broadening of $\Gamma\!\operatorname{-}\!\mathrm{LO}$ and $\Gamma\!\operatorname{-}\!\mathrm{TO}$
\begin{figure}[t!]
	\begin{center}
		\includegraphics[width = \columnwidth]{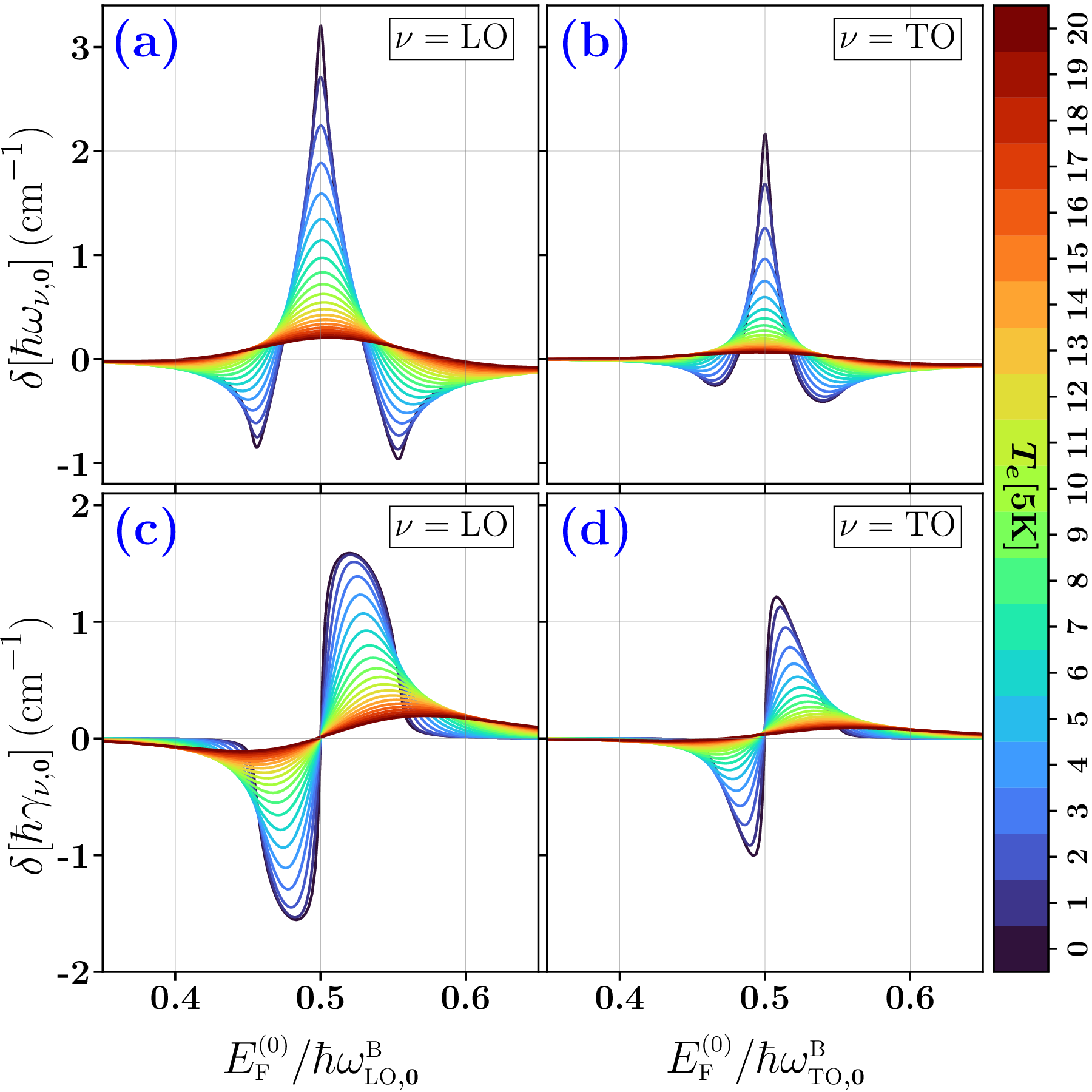}
		\caption{(Color online) Current-induced perturbation to the $\nu=\left\{\mathrm{LO},\mathrm{TO}\right\}$ modes at $\bm{q} \!=\!\bm{0}$ computed for a residual broadening of $\gamma_{\nu, \bm{0}}^{\mathrm{B}} \!=\! 8\,\mathrm{cm}^{-1}$ and a drift parameter of $\eta = 0.1$, i.e., $k_{\text{shift}}\!=\! 0.1 k_{\scriptscriptstyle{\mathrm{F}}}^{\0}$, and at temperatures ranging from $0.01$ to $100\mathrm{K}$. Panels \textbf{(a)} and \textbf{(c)} show respectively the current-induced frequency shift and the current-induced broadening of the $\Gamma\!\operatorname{-}\!\mathrm{LO}$ mode versus the normalized equilibrium-state Fermi energy; their $\Gamma\!\operatorname{-}\!\mathrm{TO}$ counterparts are shown in panels \textbf{(b)} and \textbf{(d)}. The data points corresponding to $T_{e} = 0\,\mathrm{K}$ are actually calculated for $T_{e} = 0.01\,\mathrm{K}$.}
		\label{LO_ITO_Thermal}
	\end{center}
\end{figure}
\noindent modes presented in Figs.~\ref{LO_VD} and~\ref{TO_VD} exhibit a resonant-like behavior at $E_{\scriptscriptstyle{\mathrm{F}}}^{\0} \!=\! 0.5 \, \hbar \omega^{\0}_{\Gamma} \!=\! 98\,\mathrm{meV}$, corresponding to a sample carrier density of $9.24 \times 10^{11} \mathrm{cm}^{-2}$. This behavior stems from the damping of these phonon modes due to electron-hole pair creation \cite{Yan_2007}. As it can be seen in Figs.~\ref{LO_VD} and~\ref{TO_VD}, the onset of such damping can be tuned by DC current, and in order for the current-induced perturbations to be non-negligible, the sample carrier concentration should be \textit{roughly} within the following range,
\begin{equation}\label{EF_GOLDEN_RANGE}
\frac{ \hbar \omega^{\0}_{\Gamma}}{2\left[1 - \eta\right]} \gtrsim E_{\scriptscriptstyle{\mathrm{F}}}^{\0} \gtrsim \frac{ \hbar \omega^{\0}_{\Gamma}}{2\left[1 + \eta\right]}.
\end{equation}
The expressions for the upper/lower bounds in Eq.~(\ref{EF_GOLDEN_RANGE}) can be derived at $T_{e} = 0 \, \mathrm{K}$, for a clean sample and within the low-current regime, i.e., $\eta^{2} \!\ll\! 1$, which will be discussed in Sec.~\ref{subsec:ARFACGSATEZ}. However, Eq.~(\ref{EF_GOLDEN_RANGE}) can also be applied to the cases wherein the sample temperature and residual broadening are nonzero and moderately low to obtain an estimate for the range of sample carrier densities where the current-induced perturbations are non-negligible.

As shown in Fig.~\ref{LO_ITO_Thermal}, the temperature of the electron gas adversely impacts the current-induced perturbations to the $\Gamma\!\operatorname{-}\!\mathrm{LO}$ and $\Gamma\!\operatorname{-}\!\mathrm{TO}$ modes. Since the renormalization of
\begin{figure}[t!]
	\begin{center}
		\includegraphics[width = \columnwidth]{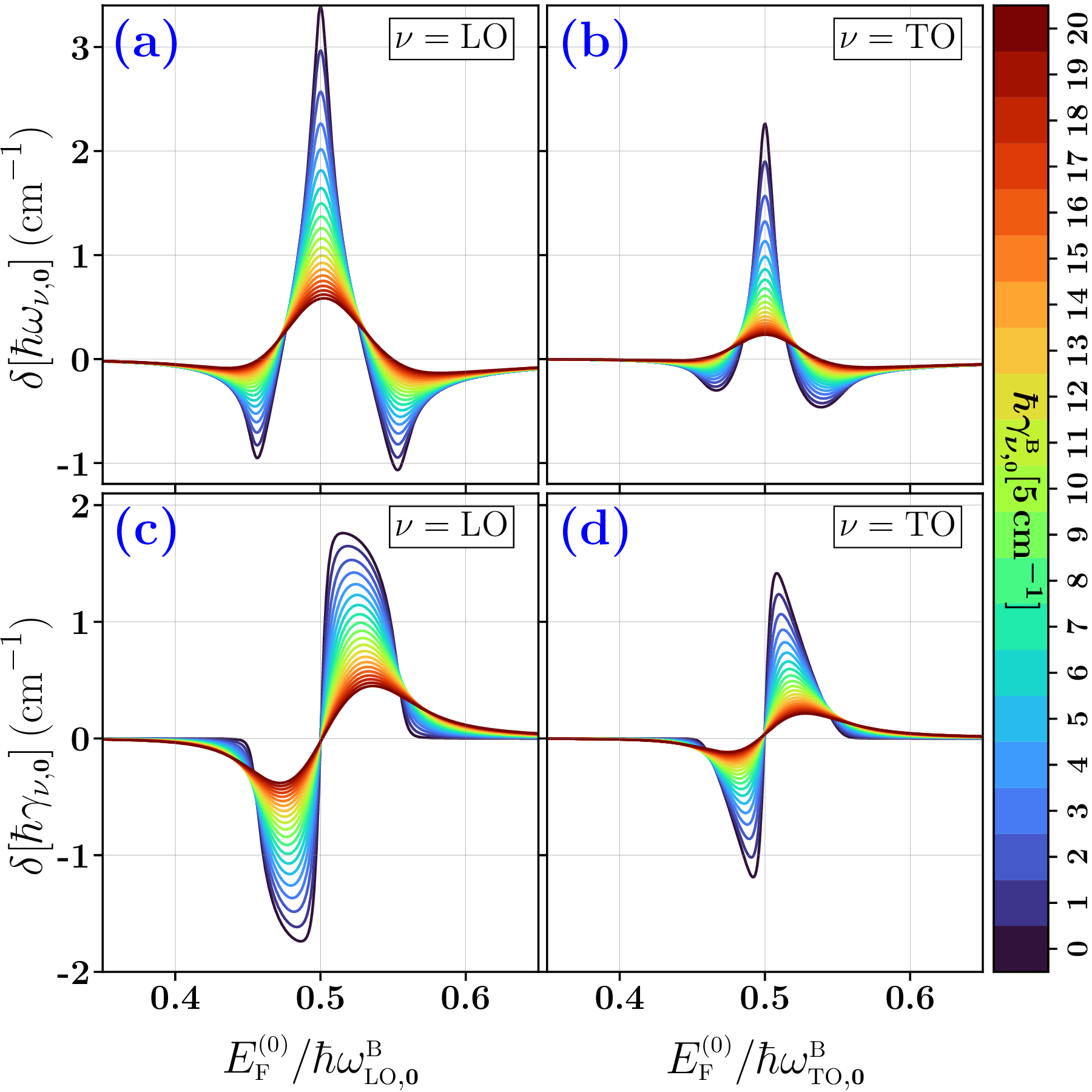}
		\caption{(Color online) Current-induced perturbation to the $\nu=\left\{\mathrm{LO},\mathrm{TO}\right\}$ modes at $\bm{q} \!=\!\bm{0}$ computed at $T_{e} = 5\mathrm{K}$, for a drift parameter of $\eta = 0.1$, i.e., $k_{\text{shift}}\!=\! 0.1 k_{\scriptscriptstyle{\mathrm{F}}}^{\0}$, and for multiple values of residual broadening, $\gamma_{\nu, \bm{0}}^{\mathrm{B}}$, ranging from $1$ to $100\,\mathrm{cm}^{-1}$. Panels \textbf{(a)} and \textbf{(c)} show respectively the current-induced frequency shift and the current-induced broadening of the $\Gamma\!\operatorname{-}\!\mathrm{LO}$ mode versus the normalized equilibrium-state Fermi energy; their $\Gamma\!\operatorname{-}\!\mathrm{TO}$ counterparts are shown in panels \textbf{(b)} and \textbf{(d)}. The data points corresponding to $\gamma_{\nu, \bm{0}}^{\mathrm{B}} = 0\,\mathrm{cm}^{-1}$ are actually calculated for $\gamma_{\nu, \bm{0}}^{\mathrm{B}} = 1\,\mathrm{cm}^{-1}$.}
		\label{LO_ITO_disorder}
	\end{center}
\end{figure}
\noindent these phonon modes is purely due to ``vertical''  inter-band electronic transitions, the current-induced perturbations to these modes is expected to vanish when the thermal smearing of the Fermi level becomes of the same order as the current-induced tilt in the Fermi level, i.e., $k_{\scriptscriptstyle{\mathrm{B}}}T_{e} \!\approx \eta |E_{\scriptscriptstyle{\mathrm{F}}}^{\0}|$. For example, for a drift parameter of $\eta \!=\! 0.1$ and a Fermi energy of $E_{\scriptscriptstyle{\mathrm{F}}}^{\0} \!=\! 98\,\mathrm{meV}$, the current-induced perturbations to the $\Gamma\!\operatorname{-}\!\mathrm{LO}$ and $\Gamma\!\operatorname{-}\!\mathrm{TO}$ modes are expected to vanish at temperatures exceeding $113\,\mathrm{K}$.

The adverse impact of the residual broadening on the current-induced perturbations to the $\Gamma\!\operatorname{-}\!\mathrm{LO}$ and $\Gamma\!\operatorname{-}\!\mathrm{TO}$ modes is shown in Fig.~\ref{LO_ITO_disorder}. Since the broadening of a phonon mode can be interpreted as the uncertainty with which the mode frequency can be determined, the shift of the Fermi disk cannot be distinguished by the ``vertical'' inter-band electronic transitions caused by the $\Gamma\!\operatorname{-}\!\mathrm{LO}$ and $\Gamma\!\operatorname{-}\!\mathrm{TO}$ modes if the mode broadening is of the same order as the current-induced tilt of the Fermi energy, i.e., $\hbar \gamma_{\nu, \bm{0}} \approx \eta |E_{\scriptscriptstyle{\mathrm{F}}}^{\0}|$. For instance, for a drift parameter of $\eta = 0.1$ and a Fermi energy of $E_{\scriptscriptstyle{\mathrm{F}}}^{\0} = 98\,\mathrm{meV}$, the current-induced perturbations to the $\Gamma\!\operatorname{-}\!\mathrm{LO}$ and $\Gamma\!\operatorname{-}\!\mathrm{TO}$ modes are expected to vanish for residual broadening values exceeding $80\,\mathrm{cm}^{-1}$.

\subsection{Analytic results for a clean sample at $T_{e}=0\,\mathrm{K}$}\label{subsec:ARFACGSATEZ}

As indicated by Fig.~\ref{FBZ} and Eq.~(\ref{SELF_ENERGY}), the computation of current-induced perturbations involves a two-dimensional integration over the reciprocal space. However, in the low-current and low-temperature limit, i.e., $\eta^{2} \ll 1$ and $k_{\scriptscriptstyle{\mathrm{B}}}T_{e} \ll \eta |E_{\scriptscriptstyle{\mathrm{F}}}^{\0}|$, both the frequency shift and broadening of the $\Gamma\!\operatorname{-}\!\mathrm{LO}$ and $\Gamma\!\operatorname{-}\!\mathrm{TO}$ modes of current-carrying graphene can be approximated by a semi-analytic formula, which involves a one-dimensional polar integral, and the instructions to utilize this semi-analytic formalism can be found at the end of Appendix~\ref{ANALYTIC_FORMULA_GAMMA_APPENDIX}.

Since the self-energy integral is calculated within the framework of $2^\mathrm{nd}$ order perturbation theory, the interaction of the $\Gamma\!\operatorname{-}\!\mathrm{LO}$ and $\Gamma\!\operatorname{-}\!\mathrm{TO}$ modes with other phonon modes, quasi-particles, impurity atoms and crystal defects can be assumed to be negligible in $1^{\mathrm{st}}$ order perturbation, a scenario referred to as the ``clean-sample'' limit \cite{Ando_2006}. If the clean-sample assumption is added to the low-temperature and low-current experimental setup, the broadening of the $\left(\nu , \bm{q} \!=\! \bm{0}\right)$ mode can be approximated by an analytic expression at $T_{e} = 0\,\mathrm{K}$. The derivation steps for this expression can be found in Appendix~\ref{ANALYTIC_FORMULA_GAMMA_APPENDIX}. The analytic expression reads
\begin{equation}\label{ANALYTIC_FORMULA_FOR_MODE_BROADENING_FINALIZED}
\hbar \gamma_{\nu,\bm{0}} \cong \frac{\sqrt{27}}{2M} \left[\frac{\hbar}{a}\right]^{2} \left[\frac{\beta}{2}\right]^{2} \left[\frac{\psi + l_{\nu} \sin{\psi} \cos{\psi}}{\pi} \right],
\end{equation}
where $\psi = \psi\!\left(E_{\scriptscriptstyle{\mathrm{F}}}^{\0}\right)$ is given by
\begin{equation}\label{THETA_DEFINITION}
\psi \equiv \left\{
\begin{array}{cc}
\!\!\! \pi &\quad |E_{\scriptscriptstyle{\mathrm{F}}}^{\0}| \leq E_{\scriptscriptstyle{\mathrm{F}}}^{\textsc{\tiny{(}}\text{\scalebox{0.70}{$+$}}\textsc{\tiny{)}}}
\\[1.0ex]
\!\!\! \arccos{\!\left[B\right]} & \quad  E_{\scriptscriptstyle{\mathrm{F}}}^{\textsc{\tiny{(}}\text{\scalebox{0.70}{$+$}}\textsc{\tiny{)}}} \! \leq |E_{\scriptscriptstyle{\mathrm{F}}}^{\0}| \leq E_{\scriptscriptstyle{\mathrm{F}}}^{\textsc{\tiny{(}}\text{\scalebox{0.70}{$-$}}\textsc{\tiny{)}}}
\\[1.0ex]
\!\!\! 0 & \quad  |E_{\scriptscriptstyle{\mathrm{F}}}^{\0}| \geq E_{\scriptscriptstyle{\mathrm{F}}}^{\textsc{\tiny{(}}\text{\scalebox{0.70}{$-$}}\textsc{\tiny{)}}}
\end{array}
\right. ,
\end{equation}
and $B = B\!\left(E_{\scriptscriptstyle{\mathrm{F}}}^{\0}\right)$ is defined as
\begin{equation}\label{THETA_DEFINITION_B}
B = \frac{|E_{\scriptscriptstyle{\mathrm{F}}}^{\0}| \left( E_{\scriptscriptstyle{\mathrm{F}}}^{\textsc{\tiny{(}}\text{\scalebox{0.70}{$-$}}\textsc{\tiny{)}}} + E_{\scriptscriptstyle{\mathrm{F}}}^{\textsc{\tiny{(}}\text{\scalebox{0.70}{$+$}}\textsc{\tiny{)}}}\right) - 2E_{\scriptscriptstyle{\mathrm{F}}}^{\textsc{\tiny{(}}\text{\scalebox{0.70}{$-$}}\textsc{\tiny{)}}} E_{\scriptscriptstyle{\mathrm{F}}}^{\textsc{\tiny{(}}\text{\scalebox{0.70}{$+$}}\textsc{\tiny{)}}}}{|E_{\scriptscriptstyle{\mathrm{F}}}^{\0}|\left(E_{\scriptscriptstyle{\mathrm{F}}}^{\textsc{\tiny{(}}\text{\scalebox{0.70}{$-$}}\textsc{\tiny{)}}} - E_{\scriptscriptstyle{\mathrm{F}}}^{\textsc{\tiny{(}}\text{\scalebox{0.70}{$+$}}\textsc{\tiny{)}}}\right)},
\end{equation}
with $E_{\scriptscriptstyle{\mathrm{F}}}^{\textsc{\tiny{(}}\text{\scalebox{0.8}{$\pm$}}\textsc{\tiny{)}}}$ being
\begin{equation}\label{EF_PM}
E_{\scriptscriptstyle{\mathrm{F}}}^{\textsc{\tiny{(}}\text{\scalebox{0.8}{$\pm$}}\textsc{\tiny{)}}} = \frac{\hbar \omega^{\0}_{\Gamma}}{2\left[1 \pm \eta\right]}.
\end{equation}
The current-induced frequency shift and broadening calculated from the semi-analytic formalism are presented in Fig.~\ref{LO_ITO_disorder_semi_analytic}, and it can be seen that the semi-analytic values for mode broadening approach those given by the analytic formalism as the residual broadening decreases.

\subsection{The impact of charge density inhomogeneity}\label{subsec:TIOCI}

As can be seen in Figs.~\ref{LO_VD}--\ref{LO_ITO_disorder}, when the equilibrium-state Fermi energy of the sample is set to be around $E_{\scriptscriptstyle{\mathrm{F}}}^{\0} = \hbar \omega^{\0}_{\Gamma}/2 \cong 98\,\mathrm{meV}$, which corresponds to a carrier density of $n_{s} \cong 9.24\times 10^{11}\mathrm{cm}^{-2}$, the current-induced perturbations to the $\Gamma\!\operatorname{-}\!\mathrm{LO}$ and $\Gamma\!\operatorname{-}\!\mathrm{TO}$ modes become sensitive to the carrier concentration of the sample. At this carrier concentration, $\delta \gamma_{\nu, \bm{0}}$ vanishes while $\delta \omega_{\nu, \bm{0}}$ is maximal for both $\nu = \left\{\mathrm{LO} , \mathrm{TO}\right\}$ modes, provided that the carrier density is uniform over the area of the graphene sample that
\begin{figure}[t!]
	\begin{center}
		\includegraphics[width = \columnwidth]{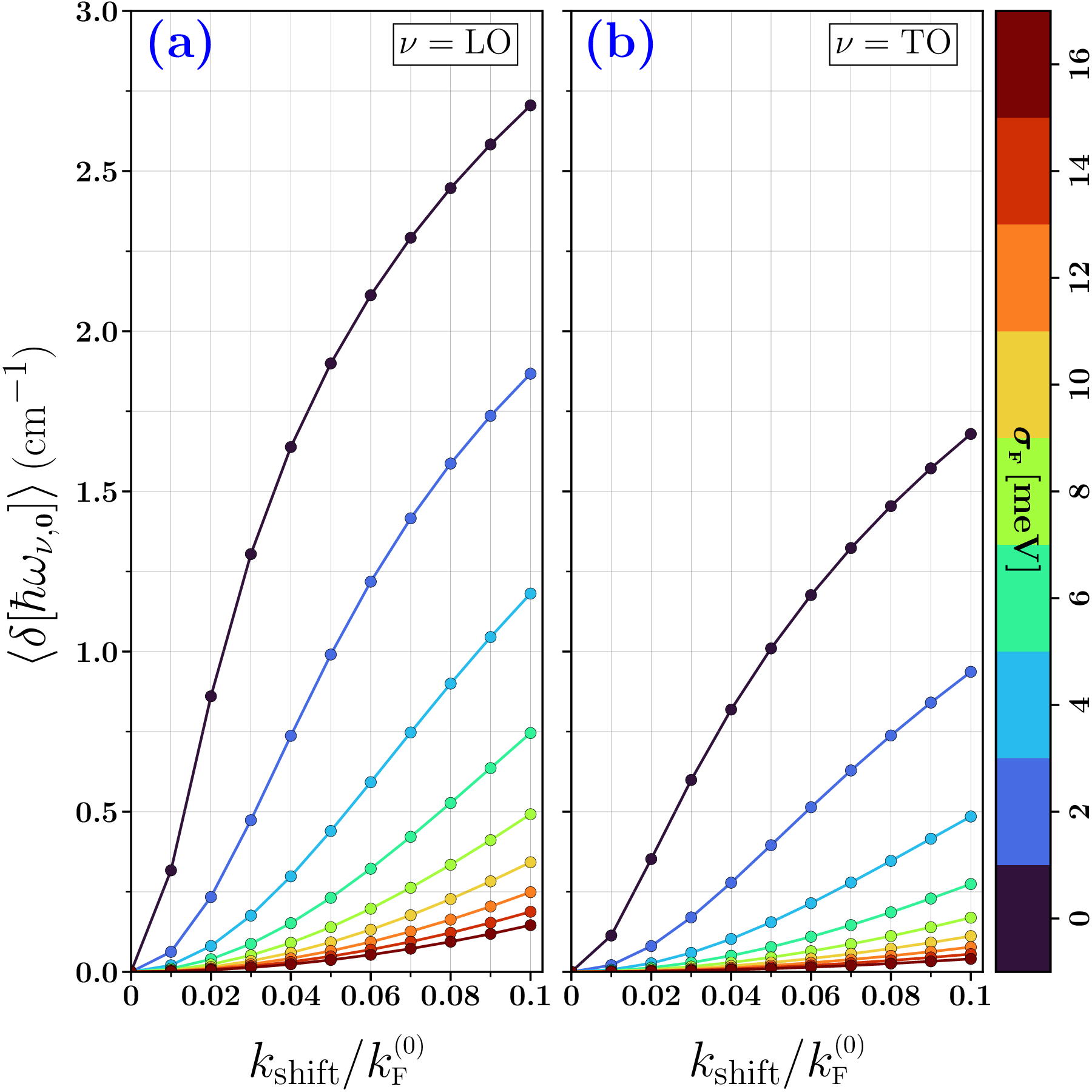}
		\caption{(Color online) Spatial averages of the current-induced frequency shift computed for the \textbf{(a)} $\Gamma\!\operatorname{-}\!\mathrm{LO}$ and \textbf{(b)} $\Gamma\!\operatorname{-}\!\mathrm{TO}$ modes versus the relative shift of the Fermi disk (or equivalently, versus the DC current). The averaging of the frequency shifts is performed assuming an average Fermi energy of $\langle E_{\scriptscriptstyle{\mathrm{F}}}^{\0} \rangle = 0.5 \hbar \omega^{\0}_{\Gamma} \cong 98\,\mathrm{meV}$ (corresponding to $\langle n_{s} \rangle \cong 9.24\times 10^{11}\mathrm{cm}^{-2}$) for $9$ values of Fermi energy variance. Prior to averaging, the current-induced perturbations were calculated at $T_{e} \!=\! 5\mathrm{K}$ for a residual broadening of $\gamma_{\nu, \bm{0}}^{\mathrm{B}} \!=\! 8\,\mathrm{cm}^{-1}$ for both modes. The dependence of the raw (unaveraged) data on $E_{\scriptscriptstyle{\mathrm{F}}}^{\0}$ is shown in the insets of Figs.~\ref{LO_VD}-\textbf{(a)} and ~\ref{TO_VD}-\textbf{(a)}.}
		\label{CFK_CURVES}
	\end{center}
\end{figure}
\noindent is irradiated by Raman laser. However, the carrier density throughout a typical graphene sample undergoes local fluctuations which are manifested as electron and hole puddles \cite{Martin_2008,Zhang_2009,Xue_2011,Decker_2011,Yankowitz_2019}. On the other hand, the laser light with which Raman spectroscopy is performed can be focused down to a spot of $1\mu\mathrm{m}$ in diameter \cite{Malard_2009,Freitag_2009,Yin_2014,Neumann_2015,Stubrov_2017}. Since the size of the charge puddles can be as small as $10\mathrm{nm}$ \cite{Martin_2008,Zhang_2009,Xue_2011,Decker_2011,Yankowitz_2019}, and therefore much smaller than the laser spot size, the impact of the carrier density fluctuations on Raman measurements cannot be neglected. The impact of charge non-uniformity on the self-energy corrections, $\Pi_{\nu,\bm{q}}$, can be quantified by performing a spatial averaging of the self-energy over the sample points under the laser spot \cite{Yan_2007,YAN200739}. This can be achieved by using a Gaussian distribution to describe the statistics of the carrier density fluctuations \cite{Martin_2008},
\begin{equation}\label{GAUSSIAN_DISTRIBUTION_OF_CHARGE_DENSITY}
p\left(E_{\scriptscriptstyle{\mathrm{F}}}^{\0}\right) = \frac{1}{\sqrt{2 \pi \sigma_{\scriptscriptstyle{\mathrm{F}}}^{2}}} \, \mathrm{exp}\!\left[-\frac{1}{2} \left(\frac{E_{\scriptscriptstyle{\mathrm{F}}}^{\0} - \langle E_{\scriptscriptstyle{\mathrm{F}}}^{\0} \rangle}{\sigma_{\scriptscriptstyle{\mathrm{F}}}}\right)^{\!2}\right],
\end{equation}
with $\langle E_{\scriptscriptstyle{\mathrm{F}}}^{\0} \rangle$ and $\sigma_{\scriptscriptstyle{\mathrm{F}}}$ being respectively the spatial average and the variance of the Fermi energy \footnote{$\langle \ldots \rangle$ indicates the spatial averaging over the sample points under the laser spot.}. The latter reflects
\begin{figure}[t!]
	\begin{center}
		\includegraphics[width = \columnwidth]{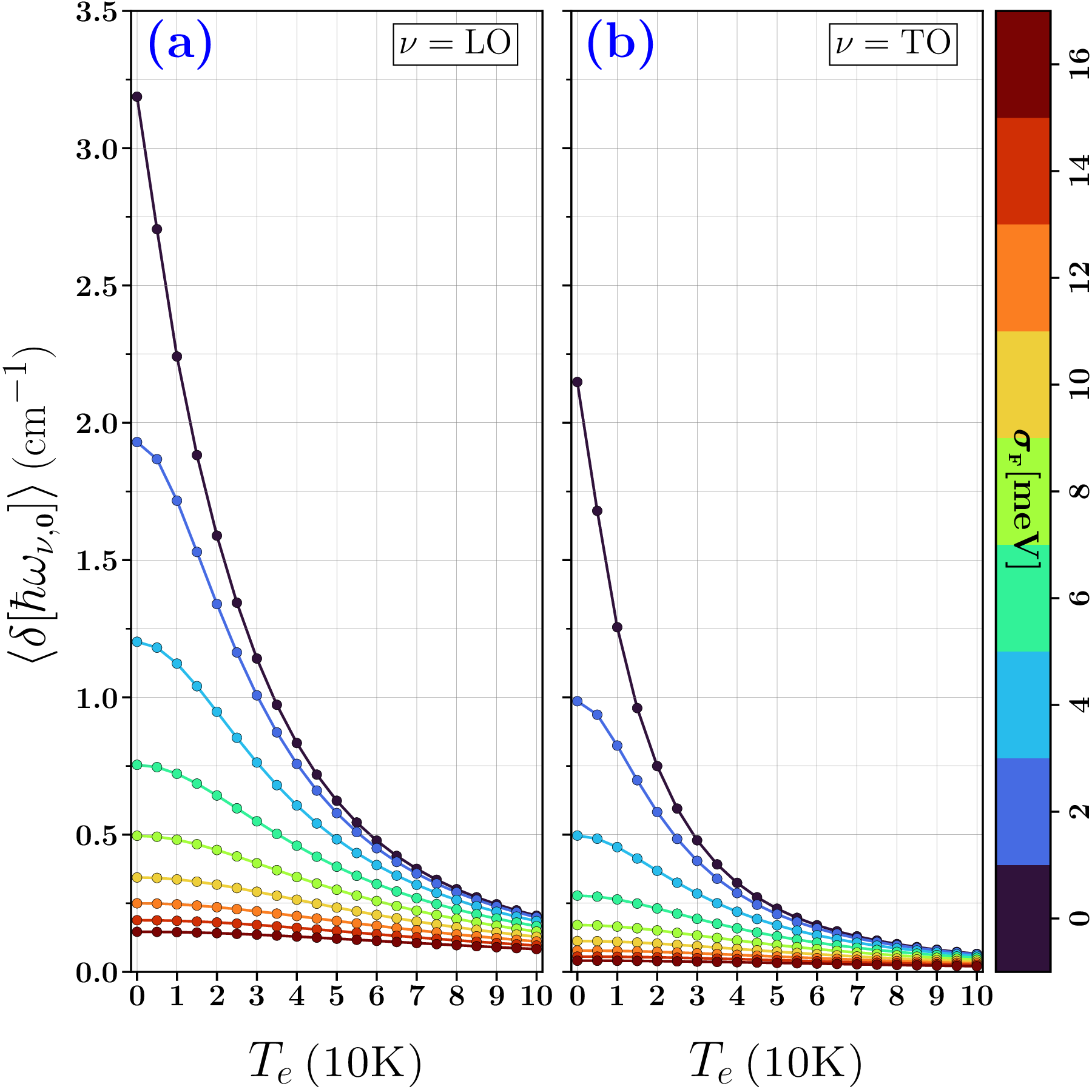}
		\caption{(Color online) Spatial averages of the current-induced frequency shift computed for the \textbf{(a)} $\Gamma\!\operatorname{-}\!\mathrm{LO}$ and \textbf{(b)} $\Gamma\!\operatorname{-}\!\mathrm{TO}$ modes versus temperature. The averaging of the frequency shifts is performed assuming an average Fermi energy of $\langle E_{\scriptscriptstyle{\mathrm{F}}}^{\0} \rangle = 0.5\hbar \omega^{\0}_{\Gamma} \cong 98\,\mathrm{meV}$ (corresponding to $\langle n_{s} \rangle \cong 9.24\times 10^{11}\mathrm{cm}^{-2}$) for $9$ values of Fermi energy variance. Prior to averaging, the current-induced perturbations were calculated for a drift parameter of $k_{\text{shift}}\!=\! 0.1 k_{\scriptscriptstyle{\mathrm{F}}}^{\0}$ and a residual broadening of $\gamma_{\nu, \bm{0}}^{\mathrm{B}} \!=\! 8\,\mathrm{cm}^{-1}$ for both modes. The dependence of the raw (unaveraged) data on $E_{\scriptscriptstyle{\mathrm{F}}}^{\0}$ is shown in panels \textbf{(a)} and \textbf{(b)} of Fig.~\ref{LO_ITO_Thermal}.}
		\label{CFT_CURVES}
	\end{center}
\end{figure}
\noindent the severity of charge nonuniformity due to the electron/hole puddles \cite{Martin_2008} which can be expressed in terms of the average and the variance of carrier density, that are denoted by respectively $\langle n_{s} \rangle$ and $\mathrm{Var}\!\left[n_{s}\right]$, as follows
\begin{equation}\label{VARIANCE}
\sigma_{\scriptscriptstyle{\mathrm{F}}} = \langle E_{\scriptscriptstyle{\mathrm{F}}}^{\0} \rangle \frac{\mathrm{Var}\!\left[n_{s}\right]}{2 \langle n_{s} \rangle}.
\end{equation}
Same averaging approach can be adopted to incorporate the impact of carrier density fluctuations on the current-induced perturbations, $\delta\Pi_{\nu,\bm{q}}$. That is
\begin{equation}\label{SPATIAL_AVERAGING}
\langle \delta\Pi_{\nu,\bm{q}} \rangle = \int_{-\infty}^{\infty}{ p\left(E_{\scriptscriptstyle{\mathrm{F}}}^{\0}\right)\delta\Pi_{\nu,\bm{q}} \, \mathrm{d}E_{\scriptscriptstyle{\mathrm{F}}}^{\0}}.
\end{equation}
The carrier density fluctuations are determined by multiple factors such as the type of substrate. Refs.~\onlinecite{Yan_2007,YAN200739} report a fluctuation of $\pm 3 \times 10^{11} \, \mathrm{cm}^{-2}$ in carrier concentration, and for our case, we adopt this value for $\mathrm{Var}\!\left[n_{s}\right]$. Assuming the gate voltage to be set to a value that corresponds to an average Fermi energy of $\langle E_{\scriptscriptstyle{\mathrm{F}}}^{\0} \rangle = \hbar \omega^{\0}_{\Gamma}/2$, such charge density variation translates into a variance in Fermi energy which is determined by Eq.~(\ref{VARIANCE}) to be $\sigma_{\scriptscriptstyle{\mathrm{F}}} \cong 0.16 \langle E_{\scriptscriptstyle{\mathrm{F}}}^{\0} \rangle \cong 16\,\mathrm{meV}$. As can be seen in Figs.~\ref{LO_VD}--\ref{LO_ITO_disorder},
\begin{figure}[t!]
	\begin{center}
		\includegraphics[width = \columnwidth]{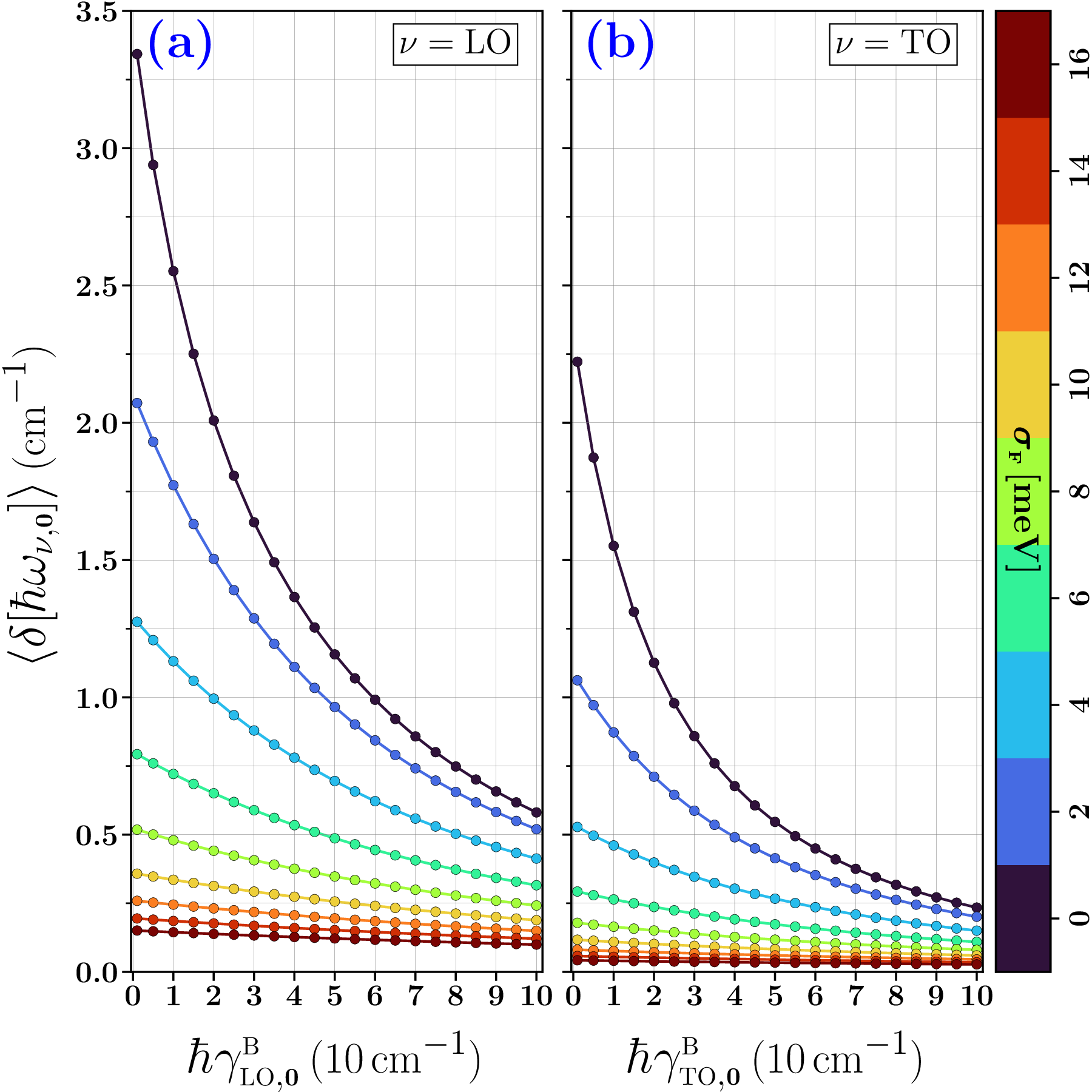}
		\caption{(Color online) Spatial averages of the current-induced frequency shift computed for the \textbf{(a)} $\Gamma\!\operatorname{-}\!\mathrm{LO}$ and \textbf{(b)} $\Gamma\!\operatorname{-}\!\mathrm{TO}$ modes versus the residual phonon mode broadening. The averaging of the frequency shifts is performed assuming an average Fermi energy of $\langle E_{\scriptscriptstyle{\mathrm{F}}}^{\0} \rangle = 0.5\hbar \omega^{\0}_{\Gamma} \cong 98\,\mathrm{meV}$ (corresponding to $\langle n_{s} \rangle \cong 9.24\times 10^{11}\mathrm{cm}^{-2}$) for $9$ values of Fermi energy variance. Prior to averaging, the current-induced perturbations were calculated at $T_{e} \!=\! 5\mathrm{K}$ for a drift parameter of $k_{\text{shift}}\!=\! 0.1 k_{\scriptscriptstyle{\mathrm{F}}}^{\0}$. The dependence of the raw (unaveraged) data on $E_{\scriptscriptstyle{\mathrm{F}}}^{\0}$ is shown in panels \textbf{(a)} and \textbf{(b)} of Fig.~\ref{LO_ITO_disorder}.}
		\label{CFD_CURVES}
	\end{center}
\end{figure}
\noindent the current-induced broadening is roughly an odd function of $E_{\scriptscriptstyle{\mathrm{F}}}^{\0} - (\hbar \omega^{\0}_{\Gamma}/2)$ suggesting that the spatial average of the current-induced broadening is expected to vanish, and consistently, the computed values for $\langle\delta [\hbar\gamma_{\nu, \bm{0}}]\rangle$ did not exceed $0.1\,\mathrm{cm}^{-1}$ even for a clean sample at $T_{e} = 5\mathrm{K}$. The results for the spatial averages of the current-induced frequency shifts are presented in Figs.~\ref{CFK_CURVES}--\ref{CFD_CURVES}.

It is worth pointing out that scanning probe micro­scopy measurements indicate that charge inhomogeneity in a substrate-supported graphene sample is mainly due to the charged impurities embedded in its $\mathrm{SiO}_{2}$ substrate \cite{Martin_2008,Zhang_2009,Deshpande_2009}. In Refs.~\onlinecite{Xue_2011,Decker_2011}, charge density fluctuations in graphene were measured for two different cases using scanning tunneling spectroscopy (STS), and the comparison between the case where graphene is directly placed on an amorphous $\mathrm{SiO}_{2}$ substrate (i.e., $\mathrm{graphene}$/$\mathrm{SiO}_{2}$) and the case where a $20\,\mathrm{nm}$-thick layer of crystalline hexagonal boron nitride ($\mathrm{hBN}$) separates graphene from the $\mathrm{SiO}_{2}$ substrate (i.e., $\mathrm{graphene}$/$\mathrm{hBN}$/$\mathrm{SiO}_{2}$) indicates that the charge inhomogeneity variance in graphene can be considerably suppressed by the placement of $\mathrm{hBN}$ layer. Moreover, the STS measurements reported in Ref.~\onlinecite{Yankowitz_2019} indicate that the separation of the $\mathrm{hBN}$ and $\mathrm{SiO}_{2}$ layers with a graphite crystal (i.e., $\mathrm{graphene}$/$\mathrm{hBN}$/$\mathrm{graphite}$/$\mathrm{SiO}_{2}$) substantially enhances the charge uniformity in graphene comparing to the $\mathrm{graphene}$/$\mathrm{hBN}$/$\mathrm{SiO}_{2}$ stacked structure.

%%%%%%%%%%%%%%%%%%%%%%%%%%%%%%%%%%%%%%%%%%%%%%%%%%%%%%%%%%%%%
%SECTION V: Current-induced perturbation to the in-plane transverse optical phonon modes at the FBZ corners
%%%%%%%%%%%%%%%%%%%%%%%%%%%%%%%%%%%%%%%%%%%%%%%%%%%%%%%%%%%%%
\section{Current-induced perturbation to the in-plane transverse optical phonon modes at the FBZ corners}\label{sec:OPT_K}
\subsection{Scattering amplitude formalism}\label{subsec:FOR_V}

Due to their large momentum, the $\mathrm{K}_{j}\!\operatorname{-}\!\mathrm{TO}$ phonon modes ($j=1,\ldots,6$) are capable of causing electronic transitions from an eigen-state $\ket{\bm{k},s}$ of one valley into the $\ket{\bm{k}+\bm{q} , s^{\prime}}$ eigen-state of the adjacent valley, i.e.,
\begin{equation}\label{VALLEY_DISTANCE}
q \sim \left|\mathbf{K}_{j}\right| = \frac{4\pi}{3\sqrt{3}a} \qquad {;}j=1,\ldots,6.
\end{equation}
Assuming the mode momentum vector, $\bm{q}$, to connect $\Gamma$ to the points in the vicinity of $\mathbf{K}_{j}$ (see Fig.~\ref{FBZ}), the scattering amplitude of the inter-valley processes involving such $\left(\mathrm{TO} , \bm{q}\right)$ modes is given by \cite{Piscanec_2004,Lazzeri_2005,Butscher_2007,Sasaki_2012,Tomadin_2013,Park_2014,Sohier_2014,Sohier_2015},
\begin{equation}\label{ELPH_SA_K}
\mathrm{F}^{[\mathrm{TO}]}_{s,s^{\prime}}\! \left(\bm{k},\bm{q}\right) = \kappa^{2}_{\mathrm{K}} \! \left[\frac{1 - ss^{\prime} \cos{\left(\theta_{\bm{k} + \bm{q}^{\prime}} - \theta_{\bm{k}} \right)}}{2}\right],
\end{equation}
where $\theta_{\bm{k}} \!\equiv\! \angle{\left(\bm{k},\hat{\bm{x}}\right)}$ and $\theta_{\bm{k}+ \bm{q}^{\prime}} \!\equiv\! \angle{\left( \bm{k} + \bm{q} - \mathbf{K}_{j} , \hat{\bm{x}} \right)}$ are the angles defined for the crystal momentum vectors $\bm{k}$ and $\bm{k} + \bm{q} - \mathbf{K}_{j}$ that are both measured with respect to the same given FBZ corner. The coupling parameter $\kappa_{\mathrm{K}}$ obtained from the TB model is as follows \cite{Park_2014,Sohier_2014,Sohier_2015},
\begin{equation}\label{COUPLING_PARAMETER_TO_K}
\kappa_{\mathrm{K}} = \frac{3t\beta}{a} \sqrt{\frac{\hbar}{2 M \omega^{\0}_{\mathrm{K}}}} \cong 0.37 \, \mathrm{eV},
\end{equation}
with $\omega^{\0}_{\mathrm{K}} = \omega^{\0}_{\mathrm{TO} , \mathbf{K}}$ being the $\mathrm{TO}$ mode frequency at $\bm{q} = \mathbf{K}_{1},\ldots,\mathbf{K}_{6}$ in the absence of DC current. Multiple measured values
\begin{figure}[t!]
	\begin{center}
		\includegraphics[width = \columnwidth]{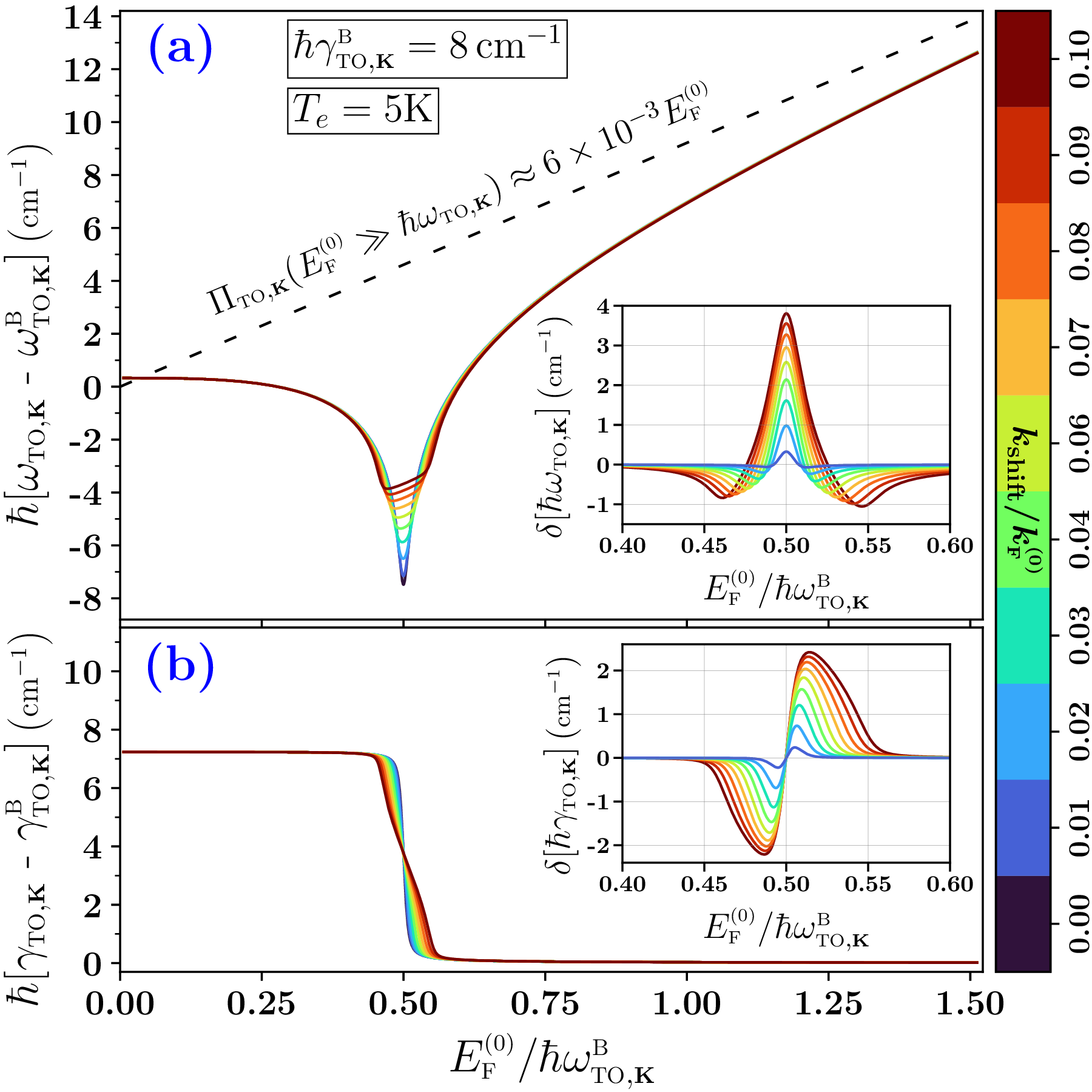}
		\caption{(Color online) The self-energy integral obtained for the $\mathrm{K}_{j}\!\operatorname{-}\!\mathrm{TO}$ mode by combining Eqs.~(\ref{SELF_ENERGY})--(\ref{THE_TB_FUNCTION}),~(\ref{Nonequilibrium_EF}),~(\ref{ELPH_SA_K}),~(\ref{COUPLING_PARAMETER_TO_K}) and~(\ref{VE_BG_K}) for Fermi disk shift values up to $0.1 k_{\scriptscriptstyle{\mathrm{F}}}^{\0}$ at $T_{e} \!=\! 5\mathrm{K}$. Panels \textbf{(a)} and \textbf{(b)} show respectively the frequency and broadening of the $\mathrm{K}_{j}\!\operatorname{-}\!\mathrm{TO}$ mode of current-carrying graphene versus the equilibrium-state Fermi energy normalized by the frequency of the $\mathrm{K}_{j}\!\operatorname{-}\!\mathrm{TO}$ mode of undoped graphene. The inset in each panel shows the current-induced perturbation to the quantity shown in its respective panel versus the normalized $E_{\scriptscriptstyle{\mathrm{F}}}^{\0}$. The Fermi energy is varied from $0$ to $250\,\mathrm{meV}$, and the current-induced perturbations exhibit a resonant-like behavior at a sample carrier density corresponding to $E_{\scriptscriptstyle{\mathrm{F}}}^{\0} = \hbar \omega^{\0}_{\mathrm{K}}/2 \cong 83\,\mathrm{meV}$.}
		\label{TO_VD_K}
	\end{center}
\end{figure}
\noindent have been reported for $\hbar\omega^{\0}_{\mathrm{K}}$, including $161.2\,\mathrm{meV}$ \cite{Piscanec_2004,Lazzeri_2005}, $149.8\,\mathrm{meV}$ \cite{Gruneis_2009_May}, $166\,\mathrm{meV}$ \cite{Gruneis_2009_August}, and $154\,\mathrm{meV}$ \cite{Popov_2010}. Since the squared scattering amplitude, and therefore the current-induced perturbation to the $\mathrm{K}_{j}\!\operatorname{-}\!\mathrm{TO}$ mode, is inversely proportional to the mode frequency, we make the conservative choice and accept the highest value as the nominal mode frequency, i.e., $\hbar\omega^{\0}_{\mathrm{K}} = 166\,\mathrm{meV}$, or equivalently $1339\,\mathrm{cm}^{-1}$.

Since the coupling of the $\mathrm{K}_{j}\!\operatorname{-}\!\mathrm{LO}$ phonon mode to the $\pi$ electron gas is estimated to be less than $1.6\%$ of its $\mathrm{TO}$ counterpart \cite{Tomadin_2013,Butscher_2007}, the impact of DC current on the $\mathrm{K}_{j}\!\operatorname{-}\!\mathrm{LO}$ modes is negligible and will not be studied here. The computation of the self-energy corrections, and therefore the current-induced perturbations to the $\mathrm{TO}$ mode exactly at $\bm{q} = \mathbf{K}_{j}$ (i.e., $\bm{q}^{\prime} = \bm{0}$), is similar to those of the $\bm{q} = \bm{0}$ modes discussed in Sec.~\ref{sec:OPT_GAM}, with few differences. First, due to the inter-valley nature of the electron-phonon scattering processes which contribute to the self-energy, the valley degeneracy does not appear in the formalism. Second, the difference between the bare mode frequencies at $\bm{q} = \mathbf{K}_{j}$ and $\bm{q} = \bm{0}$ leads to different values of the coupling constant and therefore different values of perturbations. In Ref.~\onlinecite{Araujo_2012}, the difference between these frequencies, i.e., $E_{\mathbf{K} \to \mathbf{K}^{\prime}} = \hbar\left[\omega^{\0}_{\Gamma} - \omega^{\0}_{\mathrm{K}}\right] \cong 30\,\mathrm{meV}$, is interpreted as the (phonon) energy required to translate an electron between two adjacent FBZ corners. Third, in addition to the difference in mode frequency, the electron-phonon coupling constant for the $\mathrm{TO}$ mode obtained from the $1^{\mathrm{st}}$-NN TB model at $\bm{q} = \mathbf{K}_{j}$ is greater than that of the $\mathrm{TO}$ mode at $\bm{q} = \bm{0}$ by a factor of $\sqrt{2}$. This is in agreement with the Density Functional Theory (DFT)
\begin{figure}[t!]
	\begin{center}
		\includegraphics[width = \columnwidth]{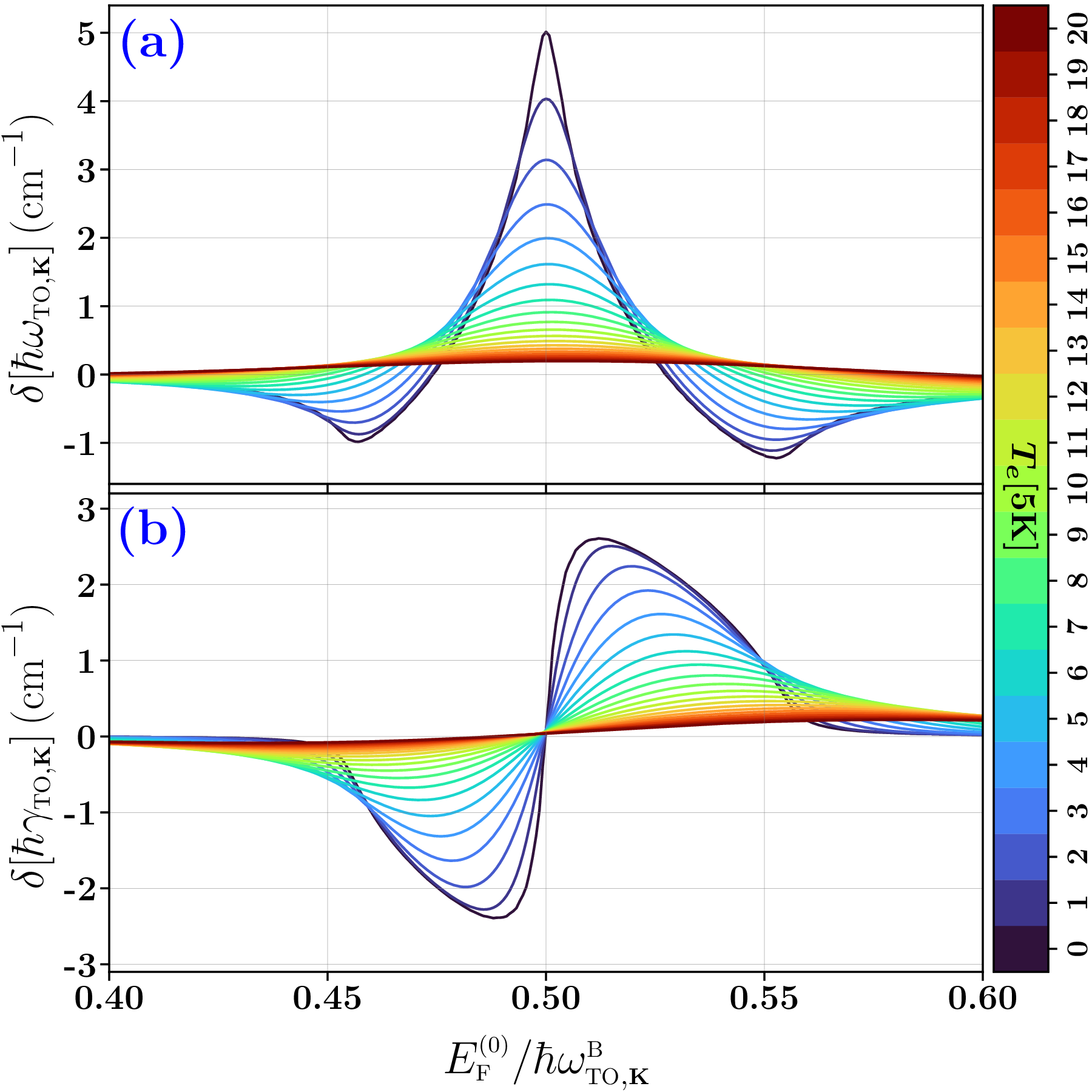}
		\caption{(Color online) Current-induced perturbation to the $\mathrm{TO}$ mode at $\bm{q} = \mathbf{K}_{1},\ldots,\mathbf{K}_{6}$ computed for a residual broadening of $\gamma_{\mathrm{TO}, \mathbf{K}}^{\mathrm{B}} \!=\! 8\,\mathrm{cm}^{-1}$ and a drift parameter of $\eta = 0.1$, i.e., $k_{\text{shift}}\!=\! 0.1 k_{\scriptscriptstyle{\mathrm{F}}}^{\0}$, and at temperatures ranging from $0.01$ to $100\mathrm{K}$. Panels \textbf{(a)} and \textbf{(b)} show respectively the current-induced frequency shift and the current-induced broadening of the $\mathrm{K}_{j}\!\operatorname{-}\!\mathrm{TO}$ mode versus the normalized equilibrium-state Fermi energy. The data points corresponding to $T_{e} = 0\,\mathrm{K}$ are actually calculated for $T_{e} = 0.01\,\mathrm{K}$.}
		\label{K_ITO_Thermal}
	\end{center}
\end{figure}
\noindent calculation result reported in Ref.~\onlinecite{Piscanec_2004}, which is expressed by $\kappa^{2}_{\mathrm{K}} \omega^{\0}_{\mathrm{K}} = 2.02 \, \kappa^{2}_{\Gamma} \omega^{\0}_{\Gamma}$. Fourth, the scattering amplitude at $\bm{q} = \mathbf{K}_{j}$ becomes independent of $\theta_{\bm{k}}$, and since the inter-band inter-valley processes are the sole contributors to the self-energy of the $\mathrm{K}_{j}\!\operatorname{-}\!\mathrm{TO}$ modes, the scattering amplitude given by Eq.~(\ref{ELPH_SA_K}) reduces to $1$.

In the case of the $\mathrm{K}_{j}\!\operatorname{-}\!\mathrm{TO}$ modes, the contribution of the virtual excitations which need to be removed from the self-energy calculations is given by,
\begin{equation}\label{VE_BG_K}
\Pi^{\scriptscriptstyle{\mathrm{V}}\scriptscriptstyle{\mathrm{E}}}_{\mathrm{TO},\mathbf{K}} = -  \left[\frac{g_{\scriptscriptstyle{\mathrm{S}}}}{2}\right] \frac{\sqrt{3}}{\pi t} \kappa^{2}_{\mathrm{K}} \left[k_{c}a\right].
\end{equation}

\begin{figure}[t!]
	\begin{center}
		\includegraphics[width = \columnwidth]{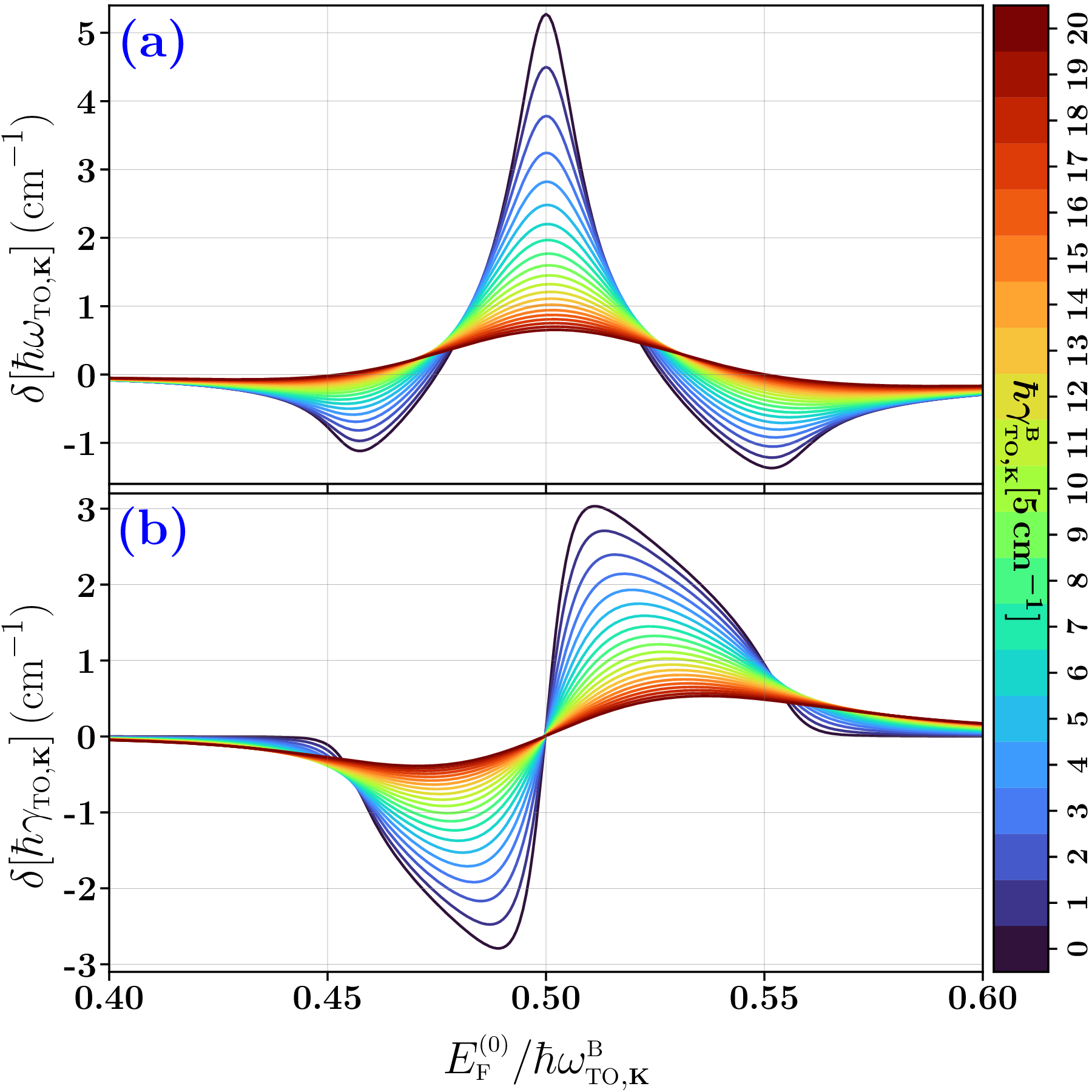}
		\caption{(Color online) Current-induced perturbation to the $\mathrm{TO}$ mode at $\bm{q} = \mathbf{K}_{1},\ldots,\mathbf{K}_{6}$ computed at $T_{e} = 5\mathrm{K}$, for a drift parameter of $\eta = 0.1$, i.e., $k_{\text{shift}}\!=\! 0.1 k_{\scriptscriptstyle{\mathrm{F}}}^{\0}$, and for multiple values of residual broadening, $\gamma_{\mathrm{TO}, \mathbf{K}}^{\mathrm{B}}$, ranging from $1$ to $100\,\mathrm{cm}^{-1}$. Panels \textbf{(a)} and \textbf{(b)} show respectively the current-induced frequency shift and the current-induced broadening of the $\mathrm{K}_{j}\!\operatorname{-}\!\mathrm{TO}$ mode versus the normalized equilibrium-state Fermi energy. The data points corresponding to $\gamma_{\mathrm{TO}, \mathbf{K}}^{\mathrm{B}} = 0\,\mathrm{cm}^{-1}$ are actually calculated for $\gamma_{\mathrm{TO}, \mathbf{K}}^{\mathrm{B}} = 1\,\mathrm{cm}^{-1}$.}
		\label{K_ITO_disorder}
	\end{center}
\end{figure}

The computed current-induced perturbations to the $\mathrm{K}_{j}\!\operatorname{-}\!\mathrm{TO}$ modes are examined here versus several parameters, including $k_{\text{shift}}$, $T_{e}$ and $\gamma_{\mathrm{TO}, \mathbf{K}}^{\mathrm{B}}$, and the numerical results have been presented in Figs.~\ref{TO_VD_K},~\ref{K_ITO_Thermal} and~\ref{K_ITO_disorder}, respectively. Since the Raman $\mathrm{G}^{\prime}$ peak involves the $\left(\mathrm{TO} , \bm{q}\right)$ modes with finite momentum, i.e., $q^{\prime} = |\bm{q} - \mathbf{K}_{j}| \lesssim a^{-1}$ \cite{Thomsen_2000,Saito_2001,Araujo_2012,Sasaki_2012,Hasdeo_2016}, the numerical results reported for the $\mathrm{K}_{j}\!\operatorname{-}\!\mathrm{TO}$ modes cannot be used to predict the impact of DC current on the $\mathrm{G}^{\prime}$ peak. Since the magnitude and direction of the reduced momentum vector, $\bf{q}^{\prime}$, of the near-zone-corner $\mathrm{TO}$ modes adds to the parameters to vary, the quantitative study of the current-induced perturbations to these modes is computationally expensive and therefore not included here.

Nonetheless, the computation of the self-energy of the near-zone-corner $\mathrm{TO}$ modes can be accelerated by converting the hexagonal FBZ domain to any of the rhombi
\noindent shown in Fig.~\ref{FBZ}, depending on which valleys are connected by $\bm{q}$. The integration can be further reduced to a single-valley integration domain by translating one of the Dirac cones. The single-valley integration routine designed to calculate the self-energy and current-induced perturbations for the $\mathrm{TO}$ modes in the vicinity of $\bm{q} = \mathbf{K}_{j}$ can also be applied to the $\bm{q} = \mathbf{K}_{j^{\prime}}$ corner by only rotating the shift direction of the Fermi disk by $\angle{\left(\mathbf{K}_{j^{\prime}},\mathbf{K}_{j}\right)}$.

The near-zone-corner $\mathrm{TO}$ modes which contribute to the $\mathrm{G}^{\prime}$ peak, are separated from the FBZ corners by a momentum proportional to the frequency of the Raman laser, i.e., $q^{\prime} \propto \omega_{L}$ \cite{Sasaki_2012,Rodriguez_2014}. Since the current-induced perturbation to the near-zone-corner $\left(\mathrm{TO}, \bm{q}^{\prime}\right)$ modes should be negligible when $q^{\prime} \!\gg\!  k_{\scriptscriptstyle{\mathrm{F}}}^{\0}$ and should be maximal when $q^{\prime} \!\sim\! k_{\scriptscriptstyle{\mathrm{F}}}^{\0}$, the impact of DC current on the $\mathrm{G}^{\prime}$ peak is expected to be maximal when $\hbar\omega_{L} \sim E_{\scriptscriptstyle{\mathrm{F}}}^{\0}$. This requires large levels of carrier concentration, i.e., $E_{\scriptscriptstyle{\mathrm{F}}}^{\0} \!\sim\! 1\mathrm{eV}$, which can be achieved using ion-gel gate dielectric \cite{Liu_2013}.

\noindent \subsection{Analytic results for a clean sample at $T_{e}=0\,\mathrm{K}$}\label{subsec:ARFACGSATEZ_K}

In the low-current and low-temperature limit, i.e., $\eta^{2} \ll 1$ and $k_{\scriptscriptstyle{\mathrm{B}}}T_{e} \ll \eta |E_{\scriptscriptstyle{\mathrm{F}}}^{\0}|$, both the frequency shift and broadening of the $\mathrm{K}_{j}\!\operatorname{-}\!\mathrm{TO}$ mode of current-carrying graphene can be approximated by a semi-analytic formula, which involves a one-dimensional polar integral, and the instructions to utilize this semi-analytic formalism can be found at the end of Appendix~\ref{ANALYTIC_FORMULA_K_APPENDIX}.

In the clean-sample, low-temperature and low-current limit, the broadening of the $\left(\mathrm{TO} , \bm{q} \!=\! \mathbf{K}_{j} \right)$ mode can be approximated by an analytic expression at $T_{e} = 0\,\mathrm{K}$. The derivation steps for this expression can be found in Appendix~\ref{ANALYTIC_FORMULA_K_APPENDIX}. The analytic expression reads
\begin{equation}\label{ANALYTIC_FORMULA_FOR_MODE_BROADENING_FINALIZED_K}
\hbar \gamma_{\mathrm{TO},\mathbf{K}} \cong \frac{\sqrt{27}}{M} \left[\frac{\hbar}{a}\right]^{2} \left[\frac{\beta}{2}\right]^{2} \left[ \frac{\psi}{\pi}\right],
\end{equation}
where $\psi = \psi\!\left(E_{\scriptscriptstyle{\mathrm{F}}}^{\0}\right)$ is given by Eqs.~(\ref{THETA_DEFINITION})--(\ref{THETA_DEFINITION_B}) in terms of
\begin{equation}\label{EF_PM_K}
E_{\scriptscriptstyle{\mathrm{F}}}^{\textsc{\tiny{(}}\text{\scalebox{0.8}{$\pm$}}\textsc{\tiny{)}}} = \frac{\hbar \omega^{\0}_{\mathrm{K}}}{2\left[1 \pm \eta\right]}.
\end{equation}
The current-induced frequency shift and broadening calculated from the semi-analytic formalism are presented in Fig.~\ref{ITO_disorder_semi_analytic_K_POINT}, and it can be seen that the semi-analytic values for mode broadening approach those given by the analytic formalism as the residual broadening decreases.

Comparing the (semi-)analytic expressions for the current-induced perturbations to the $\mathrm{K}_{j}\!\operatorname{-}\!\mathrm{TO}$ modes presented in Appendix~\ref{ANALYTIC_FORMULA_K_APPENDIX} and Eq.~(\ref{ANALYTIC_FORMULA_FOR_MODE_BROADENING_FINALIZED_K}) with their $\Gamma\!\operatorname{-}\!\mathrm{LO}$/$\Gamma\!\operatorname{-}\!\mathrm{TO}$ counterparts presented in Appendix~\ref{ANALYTIC_FORMULA_GAMMA_APPENDIX} and Eq.~(\ref{ANALYTIC_FORMULA_FOR_MODE_BROADENING_FINALIZED}), yields the following relation at $T_{e} = 0\mathrm{K}$,
\begin{equation}\label{DC_GOLDEN_RULE}
\delta \Pi_{\mathrm{TO},\mathbf{K}} = \frac{1}{g_{\scriptscriptstyle{\mathrm{V}}}} \frac{\kappa^{2}_{\mathrm{K}} \, \omega^{\0}_{\mathrm{K}}}{\kappa^{2}_{\Gamma} \, \omega^{\0}_{\Gamma}} \left[ \delta \Pi_{\mathrm{LO},\bm{0}} + \delta \Pi_{\mathrm{TO},\bm{0}}\right].
\end{equation}
The left-hand side (LHS) of Eq.~(\ref{DC_GOLDEN_RULE}) is nonzero only for carrier concentrations corresponding to $E_{\scriptscriptstyle{\mathrm{F}}}^{\0} \!\sim\! \hbar\omega^{\0}_{\mathrm{K}}/2$, while the right-hand side (RHS) is nonzero only when $E_{\scriptscriptstyle{\mathrm{F}}}^{\0} \!\sim\! \hbar\omega^{\0}_{\Gamma}/2$. Since $\omega^{\0}_{\Gamma} \!\neq\! \omega^{\0}_{\mathrm{K}}$, this equality is valid only if the $\delta \Pi$s of these three modes are computed at the same normalized frequency/broadening, i.e., the frequency/broadening of the corresponding mode expressed in units of $|E_{\scriptscriptstyle{\mathrm{F}}}^{\0}|$. The emergence of the $\omega^{\0}_{\mathrm{K}}/\omega^{\0}_{\Gamma}$ ratio in Eq.~(\ref{DC_GOLDEN_RULE}) is a result of the normalization of $k_{\scriptscriptstyle{\mathrm{F}}}^{\0}$ in each of Eqs.~(\ref{SELFENERGY_GAMMA_ILO_ITO_SIMPLIFIED}) and~(\ref{SELFENERGY_K_ITO_SIMPLIFIED}) by the frequency of the corresponding mode. Additionally, Eq.~(\ref{DC_GOLDEN_RULE}) can be generalized to finite temperatures if both sides of this equality are computed at the same normalized temperature, i.e., if the LHS is computed at $T_{e}$ the RHS should be computed at $T_{e}^{\prime}= T_{e}\, \omega^{\0}_{\mathrm{K}}/\omega^{\0}_{\Gamma}$ for the equality to hold.
%%%%%%%%%%%%%%%%%%%%%%%%%%%%%%%%%%%%%%%%%%%%%%%%%%%%%%%%%%%%%
%SECTION VI: Discussion of experimental implications
%%%%%%%%%%%%%%%%%%%%%%%%%%%%%%%%%%%%%%%%%%%%%%%%%%%%%%%%%%%%%
\section{Discussion of experimental implications}\label{sec:RAMAN}

The experimental techniques to measure the phonon dispersion in graphene include (i) inelastic neutron scattering (INS) \cite{Nicklow_1972}, (ii) high resolution electron energy-loss spectroscopy (HREELS) \cite{Siebentritt_1997,Oshima_1998,Yanagisawa_2005}, (iii) inelastic x-ray scattering (IXS) \cite{Maultzsch_2004,Mohr_2007,Gruneis_2009_August}, (iv) angle-resolved photoemission spectroscopy (ARPES) \cite{Gruneis_2009_May}, and (v) Raman spectroscopy \cite{Pocsik_1998,Tan_2002,Maultzsch_2004_b,Liu_2015}. Therefore, any of these techniques, including Raman spectroscopy, could be applied to measure the current-induced perturbations.

As was shown in Sec.~\ref{sec:OPT_GAM}, the introduction of DC electric current breaks the $\mathrm{LO}$-$\mathrm{TO}$ degeneracy at the $\Gamma$ point, and this could be manifested in the form of a splitting of the Raman $\mathrm{G}$ peak. The intensity of Raman peaks can be strictly computed using quantum mechanical perturbation theory \cite{Thomsen_2000,Basko_2008,Chen_2011,Barros_2011,Liu_2013,Hasdeo_2016,Hasdeo_dissertation}. However, it has been a common practice among experimentalists to fit Raman peaks with Lorentzians \cite{Sakata_1988,Yan_2007,Mohiuddin_2009,Berciaud_2010,Frank_2011,Yoon_2011,Huang_2014,Shioya_2014}. Therefore, to describe the current-induced $\mathrm{G}$-peak splitting, we take the simpler approach of modeling the $\mathrm{G}$-peak intensity, $I_{\mathrm{G}}\!\left(\omega_{s}\right)$, with the superposition of two Lorentzians, i.e.,
\begin{equation}\label{SPLITTED_INTENSITY}
I_{\mathrm{G}}\!\left(\omega_{s}\right) \cong \sum_{\nu \in \left\{\mathrm{LO} , \mathrm{TO}\right\}}{\frac{\mathrm{I}^{m}_{\nu} \langle \gamma_{\nu,\bm{0}}\rangle^{2}}{\left[ \omega_{s} - \langle \omega_{\nu,\bm{0}} \rangle \right]^{2} + \langle \gamma_{\nu,\bm{0}}\rangle^{2}}},
\end{equation}
with $\mathrm{I}^{m}_{\nu}$ and $\langle \omega_{\nu,\bm{0}} \rangle$/$\langle \gamma_{\nu,\bm{0}}\rangle$ being respectively the peak intensity of the Lorentzian due to the $\left(\nu , \bm{q} = \bm{0}\right)$ mode and the spatial average of mode frequency/broadening. In the absence of DC current and mechanical strain, these two modes become indistinguishable and contribute to $I_{\mathrm{G}}\!\left(\omega_{s}\right)$ identically, i.e., $\mathrm{I}^{m}_{\mathrm{LO}} = \mathrm{I}^{m}_{\mathrm{TO}}$, $\omega_{\mathrm{LO},\bm{0}} = \omega_{\mathrm{TO},\bm{0}}$ and $\gamma_{\mathrm{LO},\bm{0}} = \gamma_{\mathrm{TO},\bm{0}}$. Finally, $\omega_{s}$ denotes the Raman shift which is defined as the shift in the incident photon frequency due to the scattering processes involving the emission ($\omega_{s} < 0$) or absorption ($\omega_{s} > 0$) of phonons.

As suggested by Eq.~(\ref{SPLITTED_INTENSITY}), the contribution of each mode to the overall intensity is a Lorentzian that can be superposed onto the contribution of the other mode. However, quantum interference effects such as the dependence of peak intensity on carrier concentration \cite{Hasdeo_2016,Hasdeo_dissertation}, cannot be captured by the superposition of intensities.

\begin{figure}[t!]
	\begin{center}
		\includegraphics[width = \columnwidth]{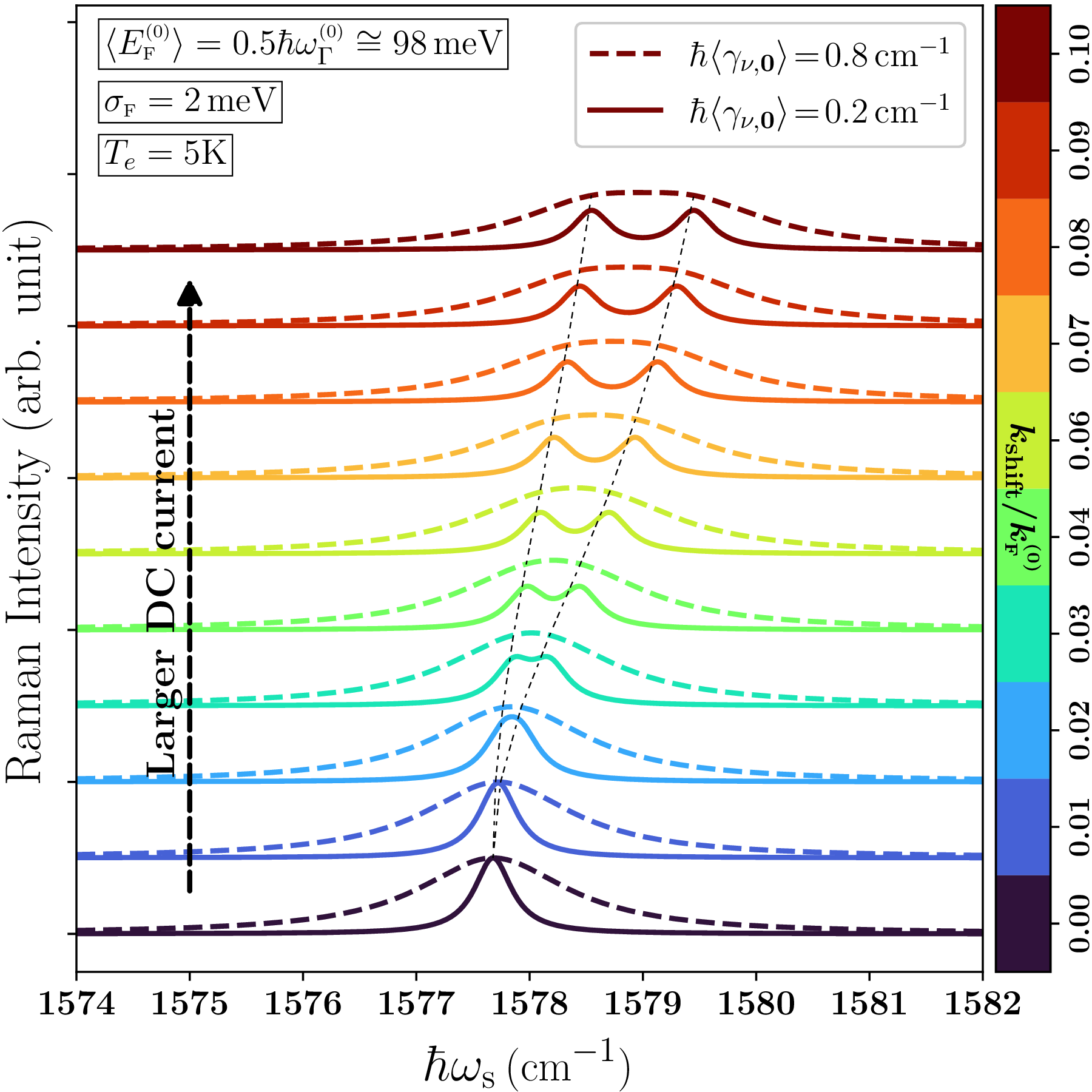}
		\caption{(Color online) Simulated Raman spectra of current-carrying graphene versus the laser frequency shift. In calculating the intensities, the spatial averages of the mode frequency shifts presented in Figs.~\ref{LO_VD} and~\ref{TO_VD} have been plugged into Eq.~(\ref{SPLITTED_INTENSITY}), and therefore the simulated Raman intensity curves represent the case where $T_{e} \!=\! 5\mathrm{K}$ and $0 \!\leq\! k_{\text{shift}} \!\leq\! 0.1 k_{\scriptscriptstyle{\mathrm{F}}}^{\0}$. An average Fermi energy of $\langle E_{\scriptscriptstyle{\mathrm{F}}}^{\0} \rangle = 0.5 \hbar \omega^{\0}_{\Gamma} \cong 98\,\mathrm{meV}$ (corresponding to a carrier concentration of $\langle n_{s} \rangle \cong 9.24\times 10^{11}\mathrm{cm}^{-2}$) and a Fermi energy variance of $\sigma_{\scriptscriptstyle{\mathrm{F}}} = 2\,\mathrm{meV}$ are assumed for the graphene sample under the laser spot. The solid and dashed curves represent the Raman intensity for the two cases wherein broadening values of respectively $\hbar\langle\gamma_{\nu, \bm{0}}\rangle \!=\! 0.2\,\mathrm{cm}^{-1}$ and $\hbar\langle\gamma_{\nu, \bm{0}}\rangle \!=\! 0.8\,\mathrm{cm}^{-1}$ have been assumed for both modes. The dashed lines trace out the peak location of the two individual $\mathrm{LO}$ and $\mathrm{TO}$ Lorentzians. For a more clear presentation of the evolution of peaks with DC current, the simulated Raman spectra have been shifted vertically.}
		\label{G_PEAK_DRIFT}
	\end{center}
\end{figure}

As it can be inferred from the phenomenological bi-Lorentzian form given by Eq.~(\ref{SPLITTED_INTENSITY}), the separation between the $\mathrm{LO}$ and $\mathrm{TO}$ peaks should be larger than the width of each of the two peaks, for the $\mathrm{G}$-peak splitting to be observable. On one hand, as shown in Figs.~\ref{CFK_CURVES}--\ref{CFD_CURVES}, $\left|\hbar\langle\gamma_{\mathrm{LO}, \bm{0}}\rangle - \hbar\langle\gamma_{\mathrm{TO}, \bm{0}}\rangle\right|$ does not exceed $1\,\mathrm{cm}^{-1}$. On the other hand, the typical value of broadening for both modes is around $10\,\mathrm{cm}^{-1}$ \cite{Yan_2007,YAN200739,Hasdeo_2016}. Therefore, this $\mathrm{G}$-peak splitting will not be observable under moderate values of current, temperature and sample disorder. To demonstrate the adverse impact of the residual mode broadening, the simulated results are presented in Fig.~\ref{G_PEAK_DRIFT} in which increasing the residual broadening from $0.2$ to $0.8\,\mathrm{cm}^{-1}$ causes the splitting to disappear. Nonetheless, it should be still possible to observe and measure the overall frequency up-shift and thickening of the $\mathrm{G}$ peak versus DC current, provided that the impact of DC current could be isolated from that of temperature.

At equilibrium, the $\mathrm{TO}$ branch in the vicinity of the $\mathrm{K}_{j}$ points can be described by a conical dispersion, with its slope being proportional to $\kappa^{2}_{\mathrm{K}}$ \cite{Piscanec_2004,Piscanec_2007}. The impact of DC current on the $\mathrm{TO}$ cones at FBZ corners can be explained as follows. On one hand, the self-energy contribution of the intra-band inter-valley transitions to the
\noindent zone-corner $\mathrm{TO}$ modes becomes non-negligible for modes of larger reduced momentum, $\bm{q}^{\prime} = \bm{q} - \mathbf{K}_{j}$ \cite{Araujo_2012,Sasaki_2012,Hasdeo_2016}. On the other hand, unlike the inter-band, the intra-band contribution to the current-induced frequency shifts are expected to persist at high temperatures. Since the impact of DC current on the \textit{intra-band} portion of the self-energy integral can be modeled by Doppler-shifted mode frequencies, the application of DC current should cause the $\mathrm{TO}$ cones to tilt at temperatures exceeding $\eta |E_{\scriptscriptstyle{\mathrm{F}}}^{\0}| / k_{\scriptscriptstyle{\mathrm{B}}}$. This tilt may be observed using IXS measurements similar to those performed in Ref.~\onlinecite{Gruneis_2009_August}.
%%%%%%%%%%%%%%%%%%%%%%%%%%%%%%%%%%%%%%%%%%%%%%%%%%%%%%%%%%%
%SECTION V: Conclusions and future works
%%%%%%%%%%%%%%%%%%%%%%%%%%%%%%%%%%%%%%%%%%%%%%%%%%%%%%%%%%%
\section{Conclusions and future works}\label{sec:S&C}

The impact of DC current on highest in-plane optical phonon modes, which include the LO mode at the FBZ center and the TO mode at the FBZ corners and center, has been studied here. The impact of several parameters such as temperature, sample disorder and carrier concentration has been explored. The current-induced perturbation to each of these modes has been shown to be nonzero only within a specific range of the sample carrier concentration. Moreover, (semi-)analytic expressions were presented which make it possible to obtain upper estimates for the current-induced perturbations without having to perform two-dimensional integration.

Due to the inter-band nature of the electronic transitions that contribute to the self-energy of $\Gamma\!\operatorname{-}\!\mathrm{LO}$ and $\Gamma\!\operatorname{-}\!\mathrm{TO}$ modes, the current-induced perturbations to these modes are sensitive to temperature and sample disorder. As a result, for moderately low values of DC current, the current-induced frequency shifts are of the same order as the Raman spectral resolution ($\sim \!1\mathrm{cm}^{-1}$). However, establishing larger DC currents in graphene sample while maintaining a relatively low temperature, i.e., $k_{\scriptscriptstyle{\mathrm{B}}}T_{e} \ll \eta |E_{\scriptscriptstyle{\mathrm{F}}}^{\0}|$, could make it possible to detect the current-induced frequency shifts.

The proposed current-induced perturbations, though moderately weak, call for the development of experimental techniques to revisit the observed evolution of Raman peaks with DC current \cite{Freitag_2009,Berciaud_2010,Yin_2014,Son_2017} to isolate the non-thermal impact of DC current on the $\mathrm{G}$ and $\mathrm{G}^{\prime}$ peaks. Clearly, the numerical results reported in this work are only valid for small DC current, i.e., $\eta^{2} \ll 1$, and in the large-current limit the NE electronic occupation should be obtained by solving the BTE. Moreover, the expressions for the coupling parameters given by Eqs.~(\ref{COUPLING_PARAMETER}) and~(\ref{COUPLING_PARAMETER_TO_K}) do not hold for extreme NE electronic occupations \cite{Hu_2022}, and therefore need to be recalculated for large values of DC current.

Raman maps of the frequency and linewidth of the $\mathrm{G}$ and $\mathrm{G}^{\prime}$ peaks in graphene have been utilized to visualize local variations in substrate, carrier concentration, mechanical stress and number of layers \cite{Zhenhua_2008,Lee_2012,Guo_2014,Neumann_2015}. Similarly, if the impact of DC current on Raman peaks can be isolated from those of temperature, carrier concentration and mechanical stress \cite{Metten_dissertation}, then local Raman measurements could be a versatile tool in determining the distribution of electric current throughout the sample, an experimental objective which has been achieved in Refs.~\onlinecite{Lee_2008,Xia_2009,Freitag_2009,Allen_2016,Bandurin_2016,Bhandari_2017,Tetienne_2017,Ella_2019,Lillie_2019,Ku_2020} owing to other physical mechanisms.

Raman intensity is the result of the constructive/destructive interference between the scattered and incident resonances \cite{Hasdeo_2016}. This leads to the dependence of Raman $\mathrm{G}$-peak intensity on carrier concentration which exhibits a peak at $2|E_{\scriptscriptstyle{\mathrm{F}}}^{\0}| = \hbar \left[\omega_{\mathrm{L}} - (\omega^{\0}_{\Gamma}/2)\right]$ \cite{Hasdeo_2016}. Moreover, in the case of uniaxially strained graphene, the Raman intensity due to the $\Gamma\!\operatorname{-}\!\mathrm{LO}$ and $\Gamma\!\operatorname{-}\!\mathrm{TO}$ modes exhibits a dependence on the angle between the laser polarization and strain axis \cite{Mohiuddin_2009,Huang_2009,Bissett_2014}. Therefore, the dependence of Raman intensity of current-carrying graphene on (i) the sample carrier concentration and (ii) the polarization of the Raman laser could be a subject of future works.

%%%%%%%%%%%%%%%%%%%%%%%%%%%%%%%%%%%%%%%%%%%%%%%%%%%%%%%%%%%
%SECTION  VII: ACKNOWLEDGEMENTS
%%%%%%%%%%%%%%%%%%%%%%%%%%%%%%%%%%%%%%%%%%%%%%%%%%%%%%%%%%%
\begin{acknowledgments}
M.S. and G.W.H. acknowledge that this work has been supported by the National Science Foundation of the United States of America under grant number EFMA-1741673. M.S. would like to thank Thibault Sohier for his suggestion to clarify the interpretation of the eigen-vectors of the current-perturbed dynamical tensor. T.S. has been supported by MICINN (Spain) under Grant No.~PID2020-113164GB-I00. J.S.G-D. acknowledges support from the National Science Foundation with a CAREER grant ECCS1749177.
\end{acknowledgments}
%%%%%%%%%%%%%%%%%%%%%%%%%%%%%%%%%%%%%%%%%%%%%%%%%%%%%%%%%%%
%SECTION  VIII: APPENDICES
%%%%%%%%%%%%%%%%%%%%%%%%%%%%%%%%%%%%%%%%%%%%%%%%%%%%%%%%%%%
\appendix
\section{Discussion on the eigen-vectors of the $\mathrm{LO}$ and $\mathrm{TO}$ modes in the presence of DC current}\label{DOMEVITPODC}

Evidently, the term $\theta_{\bm{q}}$ in Eq.~(\ref{PHI_DEFINITION}) does not vanish even for momentum vectors of extremely small magnitudes, i.e., $\theta_{q \to 0} \neq 0$. When evaluating the self-energy integral for the $\left(\mathrm{LO} , \bm{q}\right)$ and $\left(\mathrm{TO} , \bm{q}\right)$ phonon modes exactly at $\bm{q} \!=\! \bm{0}$ in the absence of DC current, the contribution of the cosine term in the scattering amplitude given by Eq.~(\ref{ELPH_SA_GAMMA}) vanishes due to the isotropic electronic occupation around the Dirac points \cite{Ando_2006}. Therefore, in the absence of the DC current, the self-energy-corrected mode frequency does not depend on the angle of the momentum vector at $\bm{q} \!=\! \bm{0}$, which is the expected behavior. However, in the presence of DC current, the anisotropic occupation of the eigen-states around the Dirac points leads to a nonzero contribution of the aforementioned cosine term which results in the dependence of mode frequency on the angle of the momentum vector, $\bm{q}$, even at $\bm{q} = \bm{0}$. This problem can be traced back to the canonical representation of phonon modes which relies on the choice of $\left\{\hat{\bm{q}}_{\parallel} , \hat{\bm{q}}_{\perp}\right\}$ unit vectors to decompose the in-plane mode displacement. These unit vectors are given by \cite{Ando_2006,Stauber_2008},
\begin{eqnarray}
\hat{\bm{q}}_{\parallel} &=& \frac{\bm{q}}{\left| \bm{q} \right|} = \hat{\bm{x}} \cos{\theta_{\bm{q}}} + \hat{\bm{y}} \sin{\theta_{\bm{q}}} \label{q_hat_para}
\\[1.0ex]
\hat{\bm{q}}_{\perp} &=& \hat{\bm{z}} \times \hat{\bm{q}}_{\parallel} = \hat{\bm{y}} \cos{\theta_{\bm{q}}}  - \hat{\bm{x}} \sin{\theta_{\bm{q}}} \label{q_hat_perp},
\end{eqnarray}
with $\hat{\bm{z}}$ being the unit vector perpendicular to the plane at which the graphene sheet is placed. In the presence of DC current, the eigen-vectors of the dynamical matrix at $\bm{q}\!=\!\bm{0}$ are expected to be the in-plane unit vectors parallel and perpendicular to the DC current. This remedy can be extended to the modes with non-zero momentum by choosing the mode eigen-vectors to be parallel and perpendicular to $\bm{q} - \bm{k}_{\text{shift}}$, which suggests that the impact of DC current on the eigen-vectors of phonon modes of larger momentum should be less significant. Therefore, in the presence of DC current we utilize the following set of orthogonal vectors as the eigen-vectors of the in-plane phonon modes of graphene,
\begin{eqnarray}
\hat{\bm{e}}_{\parallel} &=& \frac{\bm{q} - \bm{k}_{\text{shift}}}{\left| \bm{q} - \bm{k}_{\text{shift}} \right|} = \hat{\bm{x}} \cos{\Theta} + \hat{\bm{y}} \sin{\Theta} \label{e_hat_para}
\\[1.0ex]
\hat{\bm{e}}_{\!\perp} &=& \hat{\bm{z}} \times \hat{\bm{e}}_{\parallel} = \hat{\bm{y}} \cos{\Theta} - \hat{\bm{x}} \sin{\Theta} \label{e_hat_perp},
\end{eqnarray}
where $\Theta \equiv \angle{\left(\bm{q} - \bm{k}_{\text{shift}} , \hat{\bm{x}}\right)}$. Following the derivation steps presented for the equilibrium-state case in Ref.~\onlinecite{Ando_2006}, it can be readily verified that the formalism for the equilibrium-state scattering amplitude given by Eq.~(\ref{ELPH_SA_GAMMA}) can only be applied to the phonon modes of current-carrying graphene only if the angle $\theta_{\bm{q}}$ in the definition of $\phi_{\bm{k} , \bm{q}}$ is replaced with $\Theta$, i.e., $2\phi_{\bm{k} , \bm{q}} = \theta_{\bm{k} + \bm{q}} + \theta_{\bm{k}} - 2\Theta$.

This suggests that the in-plane longitudinal (transverse) modes of current-carrying graphene should be redefined to the modes with their displacement vector parallel (perpendicular) to $\bm{q} - \bm{k}_{\text{shift}}$. Another equally valid choice for the phonon mode eigen-vectors in the presence of DC current would be $\bm{q} + \bm{k}_{\text{shift}}$, i.e.,
\begin{equation}\label{QPKS}
\hat{\bm{e}}_{\parallel} = \frac{\bm{q} + \bm{k}_{\text{shift}}}{\left| \bm{q} + \bm{k}_{\text{shift}} \right|} \qquad \& \qquad \hat{\bm{q}}_{\perp} = \hat{\bm{z}} \times \hat{\bm{q}}_{\parallel}.
\end{equation}
Clearly, only one of the choices given by Eqs.~(\ref{e_hat_para}) and~(\ref{QPKS}) can be taken as the mode eigen-vector. Since the Fourier expansion in $\omega$-space has to lead to a real-valued displacement vector in time-domain, the Fourier expansion in $\bm{q}$-space should be performed over the summation variables of $\bm{q}_{-} = \bm{q} - \bm{k}_{\text{shift}}$ and $\bm{q}_{+} = \bm{q} + \bm{k}_{\text{shift}}$ for $\omega >0$ and $\omega<0$, respectively. Therefore, if the set of eigen-vectors given by Eq.~(\ref{e_hat_para}) is selected for the $\left(\nu , \bm{q}\right)$ mode of positive $\omega$, then the one given by Eq.~(\ref{QPKS}) should be reserved for the same mode with negative $\omega$. 

The results presented in this work do not depend on the generalization to the $\left|\bm{q}\right| \neq 0$ case given by Eqs.~(\ref{e_hat_para})--(\ref{e_hat_perp}) and~(\ref{QPKS}); nevertheless, these expressions are presented here as an educated guess. Even though these generalized eigen-vectors reduce to the correct result in the special cases of $\left|\bm{q}\right| = 0$ and $\left|\bm{q}\right| \gg \left|\bm{k}_{\text{shift}}\right|$, a more rigorous approach is needed to determine whether these expressions correctly describe the $\left(\mathrm{LO} , \bm{q}\right)$ and $\left(\mathrm{TO} , \bm{q}\right)$ modes of current-carrying graphene when $\left|\bm{q}\right| \sim \left|\bm{k}_{\text{shift}}\right|$. One possible approach is to directly derive the dynamical matrix of current-carrying graphene in the small-current limit, which could naturally lead us to the correct eigen-vectors for the in-plane phonon modes near FBZ center.

\section{Analytic formalism for the broadening of the $\Gamma\!\operatorname{-}\!\mathrm{LO}$ and $\Gamma\!\operatorname{-}\!\mathrm{TO}$ modes in the low-current, low-temperature and clean-sample limit}\label{ANALYTIC_FORMULA_GAMMA_APPENDIX}

At $T_{e} = 0\,\mathrm{K}$, the self-energy of the $\Gamma\!\operatorname{-}\!\mathrm{LO}$ and $\Gamma\!\operatorname{-}\!\mathrm{TO}$ modes, which is given by Eq.~(\ref{SELF_ENERGY}), can be simplified to
\begin{equation}\label{SELFENERGY_GAMMA_SIMPLIFIED}
\Pi_{\nu,\bm{0}} \cong \frac{g_{\scriptscriptstyle{\mathrm{S}}}g_{\scriptscriptstyle{\mathrm{V}}}}{A_{\scriptscriptstyle{\mathrm{F}}\scriptscriptstyle{\mathrm{B}}\scriptscriptstyle{\mathrm{Z}}}} \kappa^{2}_{\Gamma}\int_{0}^{2\pi} \! \mathrm{d}\theta_{\bm{k}} {\int_{0}^{k_{\scriptscriptstyle{\mathrm{F}}}(\theta_{\bm{k}},\theta_{d})}{\!\!\!\Lambda_{\nu,\bm{0}}\!\left(\bm{k}\right)} \, \mathrm{d}k},
\end{equation}
where $k_{\scriptscriptstyle{\mathrm{F}}}(\theta_{\bm{k}},\theta_{d})$ is described by Eq.~(\ref{Nonequilibrium_kF_SIMPLIFIED}) and $\Lambda_{\nu,\bm{0}}\!\left(\bm{k}\right)$ is
\begin{equation}\label{LAMBDA_DEFINITION}
\Lambda_{\nu,\bm{0}}\!\left(\bm{k}\right) \equiv \frac{k}{\hbar} \sum_{\alpha , \alpha^{\prime} = \pm}{\frac{0.5\left[1 - \alpha l_{\nu}\cos{\left(2\left[\theta_{\bm{k}} - \theta_{d}\right]\right)}\right]} {\left[1 - \alpha\right] v_{\scriptscriptstyle{\mathrm{F}}} k - \alpha^{\prime} \left[\omega_{\nu,\bm{0}}\!+\!i\gamma_{\nu,\bm{0}}\right]}}.
\end{equation}
As can be seen in the expression in Eq.~(\ref{LAMBDA_DEFINITION}), the two terms corresponding to $\left(\alpha , \alpha^{\prime}\right) = \left(1 , \pm 1\right)$ cancel each other, which means that the contribution of the intra-band transitions vanishes at $\bm{q} = \bm{0}$. Therefore, the expression given by Eq.~(\ref{SELFENERGY_GAMMA_SIMPLIFIED}) can be further simplified by dropping the $\alpha = 1$ term, i.e.,
\begin{equation}\label{SELFENERGY_GAMMA_ILO_ITO_SIMPLIFIED}
\left\{ \begin{matrix} \Pi_{\mathrm{LO},\bm{0}} \\[1.0ex] \Pi_{\mathrm{TO},\bm{0}} \end{matrix} \right\} \cong 2 \frac{g_{\scriptscriptstyle{\mathrm{S}}}g_{\scriptscriptstyle{\mathrm{V}}}}{A_{\scriptscriptstyle{\mathrm{F}}\scriptscriptstyle{\mathrm{B}}\scriptscriptstyle{\mathrm{Z}}}} \frac{k_{\scriptscriptstyle{\mathrm{F}}}^{\0}}{\hbar v_{\scriptscriptstyle{\mathrm{F}}}} \kappa^{2}_{\Gamma} \int_{0}^{\pi}{\!\!\mathrm{S}_{\Gamma}\!\left(\theta\right) \left\{ \begin{matrix} \cos^{2}\!{\theta} \\[1.0ex] \sin^{2}\!{\theta} \end{matrix} \right\} \mathrm{d} \theta},
\end{equation}
where $\theta \equiv \theta_{\bm{k}} - \theta_{d}$ and $\mathrm{S}_{\Gamma}\!\left(\theta\right)$ is defined as
\begin{equation}\label{K_DEFINITION}
\mathrm{S}_{\Gamma}\!\left(\theta\right) = \sum_{\alpha^{\prime} = \pm}{\int_{0}^{\tilde{k}_{\scriptscriptstyle{\mathrm{F}}}(\theta)}{\frac{\tilde{k} \, \mathrm{d}\tilde{k}} {2\tilde{k} - \alpha^{\prime} \tilde{\omega}_{\Gamma}^{c}}}},
\end{equation}
with the auxiliary variables $\tilde{k}$, $\tilde{k}_{\scriptscriptstyle{\mathrm{F}}}(\theta) $ and $\tilde{\omega}_{\Gamma}^{c}$ being
\begin{eqnarray}
&\tilde{k} \equiv \dfrac{k}{k_{\scriptscriptstyle{\mathrm{F}}}^{\0}} \label{AUX_1},
\\[1.0ex]
&\tilde{k}_{\scriptscriptstyle{\mathrm{F}}}(\theta) \equiv \dfrac{k_{\scriptscriptstyle{\mathrm{F}}}(\theta,\theta_{d} = 0)}{k_{\scriptscriptstyle{\mathrm{F}}}^{\0}} \label{AUX_2},
\\[1.0ex]
&\tilde{\omega}_{\Gamma}^{c} \equiv \dfrac{\hbar\omega^{\0}_{\Gamma}}{|E_{\scriptscriptstyle{\mathrm{F}}}^{\0}|} + i \dfrac{\hbar\gamma^{\0}_{\Gamma}}{|E_{\scriptscriptstyle{\mathrm{F}}}^{\0}|} \label{AUX_3},
\end{eqnarray}
where $\gamma^{\0}_{\Gamma} = \gamma^{\0}_{\mathrm{LO},\bm{0}} = \gamma^{\0}_{\mathrm{TO},\bm{0}}$. Note that the pre-factor ``2'' in Eq.~(\ref{SELFENERGY_GAMMA_ILO_ITO_SIMPLIFIED}) resulted from reducing the integration range from $[0, 2\pi)$ to $[0 , \pi)$ simply because the integral over $[0 , \pi)$ is equal to the integral over $[\pi , 2\pi)$. Utilizing this identity simplifies the expression in Eq.~(\ref{K_DEFINITION}),
\begin{equation}\label{Logarithmic_integral_identity}
\int{\frac{x \, \mathrm{d}x} {x + A}} = x  - A \ln{\!\left[x + A \right]} + C,
\end{equation}
with $A$ and $C$ being arbitrary constants. Therefore,
\begin{equation}\label{K_DEFINITION_SIMPLIFIED}
\mathrm{S}_{\Gamma}\!\left(\theta\right) = \left[\tilde{k} + \frac{\tilde{\omega}_{\Gamma}^{c}}{4} \ln{\! \left(\frac{2\tilde{k} - \tilde{\omega}_{\Gamma}^{c}}{2\tilde{k} + \tilde{\omega}_{\Gamma}^{c}}\right)}\right]_{0}^{\tilde{k}_{\scriptscriptstyle{\mathrm{F}}}(\theta)}.
\end{equation}
Even in the clean-sample limit, i.e., $\gamma^{\0}_{\Gamma}= 0$, the real-valued argument of $\ln{\!\left(x\right)}$ in Eq.~(\ref{K_DEFINITION_SIMPLIFIED}) can be negative, and therefore $\mathrm{K}\!\left(\theta\right)$ can be complex-valued. In that case,
\begin{equation}\label{K_DEFINITION_FURTHER_SIMPLIFIED_IMAGINARY_PART}
\mathrm{Im}\!\left[\mathrm{S}_{\Gamma}\!\left(\theta\right)\right] = -\frac{\pi \tilde{\omega}_{\Gamma}}{4} \mathrm{H}\!\left( \frac{\tilde{\omega}_{\Gamma}}{2} - \tilde{k}_{\scriptscriptstyle{\mathrm{F}}}(\theta)\right),
\end{equation}
with $\mathrm{H}\!\left(x\right)$ denoting the Heaviside step function and $\tilde{\omega}_{\Gamma} = \hbar\omega^{\0}_{\Gamma}/ |E_{\scriptscriptstyle{\mathrm{F}}}^{\0}|$. The expression given by Eq.~(\ref{K_DEFINITION_FURTHER_SIMPLIFIED_IMAGINARY_PART}) is obtained using the following identity
\begin{equation}\label{ANOTHER_IDENTITY}
\mathrm{Im}\!\left[\ln{\!\left(x\right)}\right] = \pm\pi \mathrm{H}\!\left(-x\right) \qquad {;}x\in \mathbb{R},
\end{equation}
where we accepted the ``$-$'' sign to get
\begin{figure}[t!]
	\begin{center}
		\includegraphics[width = \columnwidth]{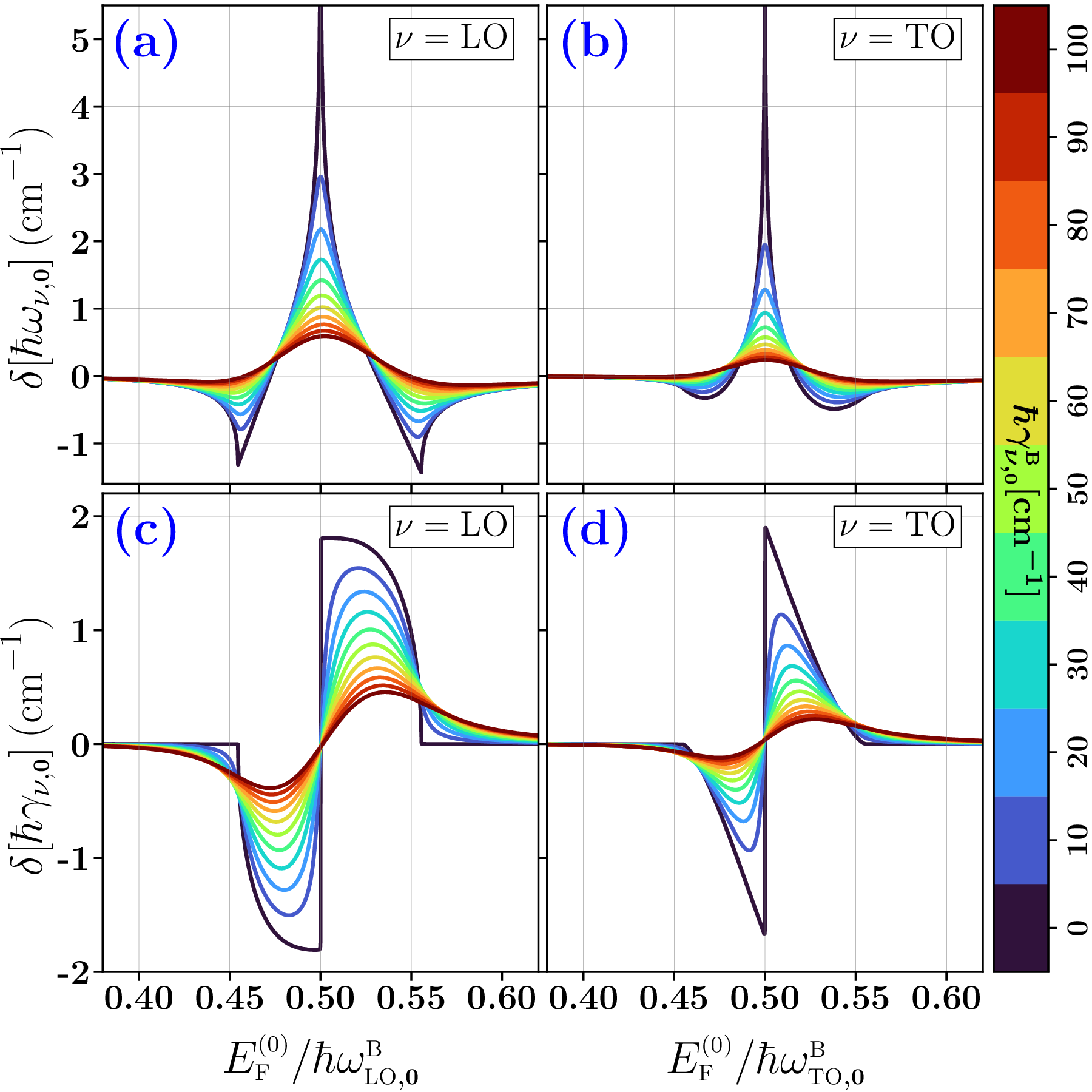}
		\caption{(Color online) Current-induced perturbation to the $\nu=\left\{\mathrm{LO},\mathrm{TO}\right\}$ modes at $\bm{q} \!=\!\bm{0}$ computed at $T_{e} = 0\mathrm{K}$, for a drift parameter of $\eta = 0.1$, i.e., $k_{\text{shift}}\!=\! 0.1 k_{\scriptscriptstyle{\mathrm{F}}}^{\0}$, and for multiple values of residual broadening, $\gamma_{\nu, \bm{0}}^{\mathrm{B}}$, ranging from $0.01$ to $100\,\mathrm{cm}^{-1}$. Panels \textbf{(a)} and \textbf{(c)} show respectively the current-induced frequency shift and the current-induced broadening of the $\Gamma\!\operatorname{-}\!\mathrm{LO}$ mode versus the normalized equilibrium-state Fermi energy; their $\Gamma\!\operatorname{-}\!\mathrm{TO}$ counterparts are shown in panels \textbf{(b)} and \textbf{(d)}. The data points corresponding to $\gamma_{\nu, \bm{0}}^{\mathrm{B}} = 0\,\mathrm{cm}^{-1}$ are actually calculated for $\gamma_{\nu, \bm{0}}^{\mathrm{B}} = 0.01\,\mathrm{cm}^{-1}$. The frequency shift and broadening of each mode are computed semi-analytically, by combining Eqs.~(\ref{SELFENERGY_GAMMA_ILO_ITO_SIMPLIFIED}) and~(\ref{K_DEFINITION_SIMPLIFIED}). The current-induced perturbations are simply computed by subtracting the self-energy values in the absence of the DC current from their counterpart computed in the presence of the DC current. The semi-analytic values for the mode broadening approach the analytic values given by Eqs.~(\ref{ANALYTIC_FORMULA_FOR_MODE_BROADENING_FINALIZED})--(\ref{EF_PM}) as the residual mode broadening becomes vanishingly small.}
		\label{LO_ITO_disorder_semi_analytic}
	\end{center}
\end{figure}
\noindent a positive value for mode broadening. Combining the expressions given by Eqs.~(\ref{SELFENERGY_GAMMA_ILO_ITO_SIMPLIFIED}) and~(\ref{K_DEFINITION_FURTHER_SIMPLIFIED_IMAGINARY_PART}) yields the following expression for the broadening of the $\Gamma\!\operatorname{-}\!\mathrm{LO}$ and $\Gamma\!\operatorname{-}\!\mathrm{TO}$ modes in the presence of DC current at $T_{e}= 0\,\mathrm{K}$,
\begin{equation}\label{BROADENING_SIMPLIFIED}
\gamma_{\nu,\bm{0}} \cong \frac{g_{\scriptscriptstyle{\mathrm{S}}}g_{\scriptscriptstyle{\mathrm{V}}}}{A_{\scriptscriptstyle{\mathrm{F}}\scriptscriptstyle{\mathrm{B}}\scriptscriptstyle{\mathrm{Z}}}} \frac{\kappa^{2}_{\Gamma}}{\hbar^{2} v^{2}_{\scriptscriptstyle{\mathrm{F}}}} \frac{\pi \omega^{\0}_{\Gamma}}{4}\!\! \int_{\vartheta}^{\pi}{\left[1 + l_{\nu} \cos\!{\left(2\theta\right)} \right] \mathrm{d}\theta},
\end{equation}
where the angle $\vartheta$ is the lower limit of the range of $\theta$ values wherein the inequality of $\tilde{k}_{\scriptscriptstyle{\mathrm{F}}}(\theta) > 0.5 \tilde{\omega}_{\Gamma}$ holds. The search for this range can be performed graphically, and $\vartheta$ the angle at which the shifted Fermi circle and the non-shifted circle of radius $\frac{\tilde{\omega}_{\Gamma}}{2}$ intersect, i.e.,
\begin{equation}\label{VARTHETA_SOLUTION}
\tilde{k}_{\scriptscriptstyle{\mathrm{F}}}(\vartheta) \cong 1 + \eta \cos{\vartheta} = \tilde{\omega}_{\Gamma}/2.
\end{equation}
Obviously, solutions for $\vartheta$ exists only if $\left||\tilde{\omega}_{\Gamma} / 2| - 1\right| \leq \eta$. The explicit solutions to Eqs.~(\ref{BROADENING_SIMPLIFIED}) and ~(\ref{VARTHETA_SOLUTION}) are presented in Sec.~\ref{subsec:ARFACGSATEZ} by Eqs.~(\ref{ANALYTIC_FORMULA_FOR_MODE_BROADENING_FINALIZED})--(\ref{EF_PM}), in terms of $\psi = \pi - \vartheta$. Additionally, in the low-current and low-temperature limit the frequency-shift and broadening of the modes can be obtained by combining Eqs.~(\ref{SELFENERGY_GAMMA_ILO_ITO_SIMPLIFIED}) and~(\ref{K_DEFINITION_SIMPLIFIED}). The current-induced perturbations can simply be computed by subtracting the values obtained from the (semi-)analytic formalism for $\eta = 0$ from their $\eta \neq 0$ counterpart. This can also be achieved by changing the lower bound in the expressions given by Eq.~(\ref{K_DEFINITION_SIMPLIFIED}) from $0$ to $k_{\scriptscriptstyle{\mathrm{F}}}^{\0}$. The computed values for the current-induced frequency shift and broadening obtained from the semi-analytic formalism are presented in Fig.~(\ref{LO_ITO_disorder_semi_analytic}).

\section{Analytic formalism for the broadening of the $\mathrm{K}_{j}\!\operatorname{-}\!\mathrm{TO}$ mode in the low-current, low-temperature and clean-sample limit}\label{ANALYTIC_FORMULA_K_APPENDIX}

Considering the differences between the $\mathrm{K}_{j}\!\operatorname{-}\!\mathrm{TO}$ and $\Gamma\!\operatorname{-}\!\mathrm{TO}$ modes in terms of the self-energy calculation, which are listed in Sec.~\ref{subsec:FOR_V}, the (semi-)analytic formalism for the $\mathrm{K}_{j}\!\operatorname{-}\!\mathrm{TO}$ modes can be obtained by making a few minor changes to the formalism presented in Appendix~\ref{ANALYTIC_FORMULA_GAMMA_APPENDIX}. Starting with Eq.~(\ref{SELFENERGY_GAMMA_SIMPLIFIED}), at $T_{e} = 0\,\mathrm{K}$, the self-energy of the $\mathrm{K}_{j}\!\operatorname{-}\!\mathrm{TO}$ mode can be simplified to
\begin{equation}\label{SELFENERGY_K_SIMPLIFIED}
\Pi_{\mathrm{TO},\mathbf{K}} \cong \frac{g_{\scriptscriptstyle{\mathrm{S}}} \kappa^{2}_{\mathrm{K}} }{A_{\scriptscriptstyle{\mathrm{F}}\scriptscriptstyle{\mathrm{B}}\scriptscriptstyle{\mathrm{Z}}}} \int_{0}^{2\pi} \! \mathrm{d}\theta_{\bm{k}} {\int_{0}^{k_{\scriptscriptstyle{\mathrm{F}}}(\theta_{\bm{k}},\theta_{d})}{\!\!\!\Lambda_{\mathrm{TO},\mathbf{K}}\!\left(\bm{k}\right)} \, \mathrm{d}k},
\end{equation}
where $\Lambda_{\mathrm{TO},\mathbf{K}}\!\left(\bm{k}\right)$ is defined as
\begin{equation}\label{LAMBDA_DEFINITION_K}
\Lambda_{\mathrm{TO},\mathbf{K}}\!\left(\bm{k}\right) \equiv \frac{k}{\hbar} \sum_{\alpha^{\prime} = \pm}{\frac{1}{2v_{\scriptscriptstyle{\mathrm{F}}} k - \alpha^{\prime} \left[\omega_{\mathrm{TO},\mathbf{K}}\!+\!i\gamma_{\mathrm{TO},\mathbf{K}}\right]}}.
\end{equation}
The semi-analytic formalism is therefore given by
\begin{equation}\label{SELFENERGY_K_ITO_SIMPLIFIED}
\Pi_{\mathrm{TO},\mathbf{K}} \cong \frac{2 g_{\scriptscriptstyle{\mathrm{S}}}}{A_{\scriptscriptstyle{\mathrm{F}}\scriptscriptstyle{\mathrm{B}}\scriptscriptstyle{\mathrm{Z}}}} \frac{k_{\scriptscriptstyle{\mathrm{F}}}^{\0}}{\hbar v_{\scriptscriptstyle{\mathrm{F}}}} \kappa^{2}_{\mathrm{K}} \int_{0}^{\pi}{\mathrm{S}_{\mathrm{K}}\!\left(\theta\right)  \mathrm{d} \theta},
\end{equation}
where $\mathrm{S}_{\mathrm{K}}\!\left(\theta\right)$ is defined as
\begin{equation}\label{S_DEFINITION_SIMPLIFIED_K_POINTS}
\mathrm{S}_{\mathrm{K}}\!\left(\theta\right) = \left[\tilde{k} + \frac{\tilde{\omega}_{\mathrm{K}}^{c}}{4} \ln{\! \left(\frac{2\tilde{k} - \tilde{\omega}_{\mathrm{K}}^{c}}{2\tilde{k} + \tilde{\omega}_{\mathrm{K}}^{c}}\right)}\right]_{0}^{\tilde{k}_{\scriptscriptstyle{\mathrm{F}}}(\theta)},
\end{equation}
with $ \tilde{\omega}_{\mathrm{K}}^{c}$ being defined as
\begin{equation}\label{OMEGA_K_c_DEFINITION}
\tilde{\omega}_{\mathrm{K}}^{c} = \tilde{\omega}_{\mathrm{K}} + i \tilde{\gamma}_{\mathrm{K}} = \frac{\hbar\omega^{\0}_{\mathrm{K}}}{|E_{\scriptscriptstyle{\mathrm{F}}}^{\0}|} + i \frac{\hbar\gamma^{\0}_{\mathrm{K}}}{|E_{\scriptscriptstyle{\mathrm{F}}}^{\0}|} \quad {;} \gamma^{\0}_{\mathrm{K}} = \gamma^{\0}_{\mathrm{TO},\mathbf{K}}.
\end{equation}
In the clean-sample limit, i.e., $\gamma^{\0}_{\mathrm{K}}= 0$, we have
\begin{equation}\label{K_DEFINITION_FURTHER_SIMPLIFIED_IMAGINARY_PART_K_POINT}
\mathrm{Im}\!\left[\mathrm{S}_{\mathrm{K}}\!\left(\theta\right)\right] = -\frac{\pi \tilde{\omega}_{\mathrm{K}}}{4} \mathrm{H}\!\left( \frac{\tilde{\omega}_{\mathrm{K}}}{2} - \tilde{k}_{\scriptscriptstyle{\mathrm{F}}}(\theta)\right).
\end{equation}
Combining the expressions given by Eqs.~(\ref{SELFENERGY_K_ITO_SIMPLIFIED}) and~(\ref{K_DEFINITION_FURTHER_SIMPLIFIED_IMAGINARY_PART_K_POINT}) yields the
\begin{figure}[t!]
	\begin{center}
		\includegraphics[width = 1.00\columnwidth]{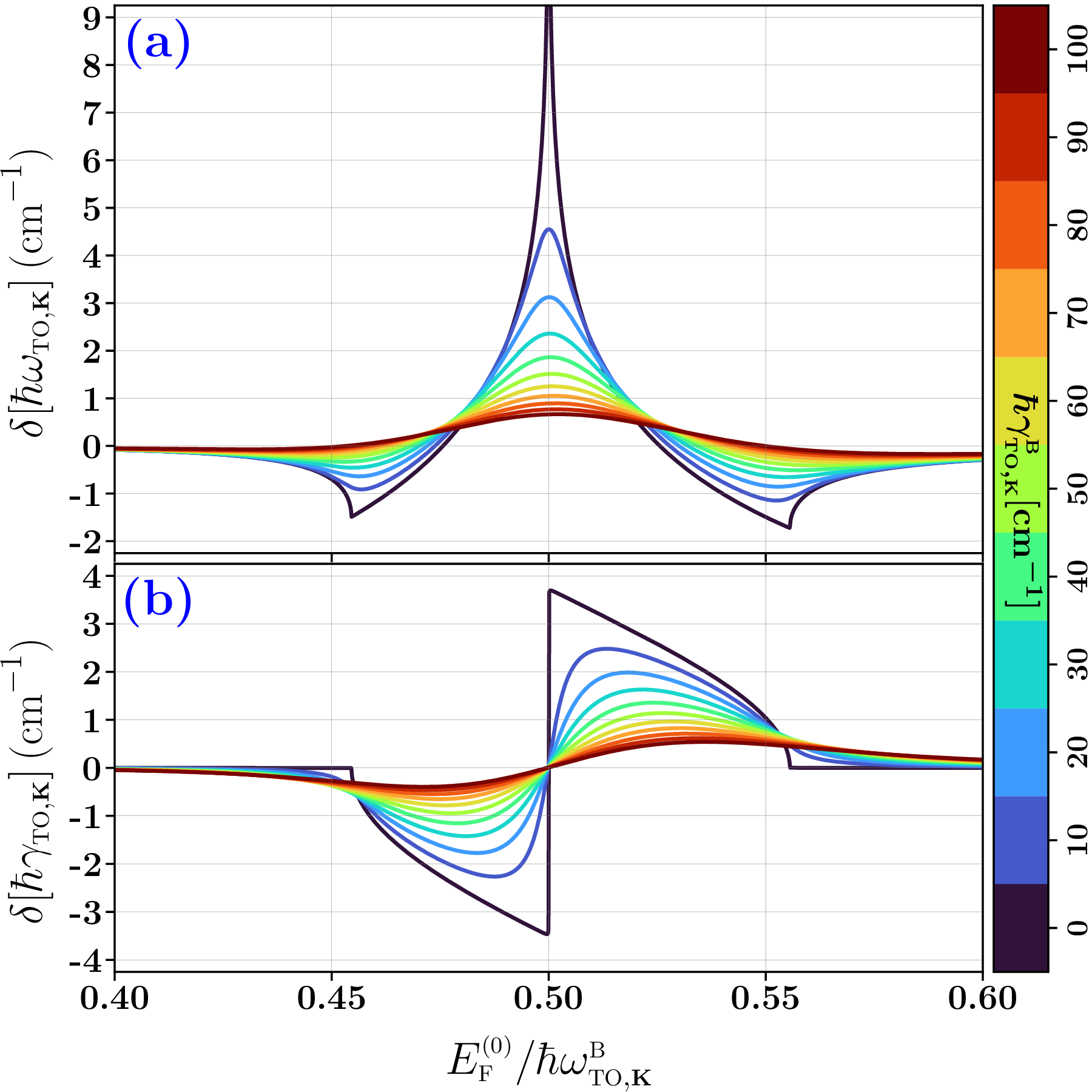}
		\caption{(Color online) Current-induced perturbation to the $\mathrm{TO}$ mode at $\bm{q} \!=\!\mathbf{K}_{j}$ computed at $T_{e} = 0\mathrm{K}$, for a drift parameter of $\eta = 0.1$, i.e., $k_{\text{shift}}\!=\! 0.1 k_{\scriptscriptstyle{\mathrm{F}}}^{\0}$, and for multiple values of residual broadening, $\gamma_{\mathrm{TO}, \mathbf{K}}^{\mathrm{B}}$, ranging from $0.01$ to $100\,\mathrm{cm}^{-1}$. Panels \textbf{(a)} and \textbf{(b)} show respectively the current-induced frequency shift and the current-induced broadening of the $\mathrm{K}_{j}\!\operatorname{-}\!\mathrm{TO}$ mode versus the normalized equilibrium-state Fermi energy. The data points corresponding to $\gamma_{\mathrm{TO}, \mathbf{K}}^{\mathrm{B}} \!=\! 0\,\mathrm{cm}^{-1}$ are actually calculated for $\gamma_{\mathrm{TO}, \mathbf{K}}^{\mathrm{B}} \!=\! 0.01\mathrm{cm}^{-1}$. The frequency shift and broadening are computed semi-analytically, by combining Eqs.~(\ref{SELFENERGY_K_ITO_SIMPLIFIED})--(\ref{S_DEFINITION_SIMPLIFIED_K_POINTS}). The current-induced perturbations are computed by subtracting the self-energy values in the absence of the DC current from their counterpart computed in the presence of the DC current. The semi-analytic values for the mode broadening approach the analytic values given by Eqs.~(\ref{ANALYTIC_FORMULA_FOR_MODE_BROADENING_FINALIZED_K})--(\ref{EF_PM_K}) upon decreasing the residual mode broadening.}
		\label{ITO_disorder_semi_analytic_K_POINT}
	\end{center}
\end{figure}
\noindent following expression for the broadening of the $\mathrm{K}_{j}\!\operatorname{-}\!\mathrm{TO}$ mode in the presence of DC current at $T_{e}= 0\,\mathrm{K}$,
\begin{equation}\label{BROADENING_SIMPLIFIED_K_POINT}
\gamma_{\mathrm{TO},\mathbf{K}} \cong \frac{2g_{\scriptscriptstyle{\mathrm{S}}}}{A_{\scriptscriptstyle{\mathrm{F}}\scriptscriptstyle{\mathrm{B}}\scriptscriptstyle{\mathrm{Z}}}} \frac{\kappa^{2}_{\mathrm{K}}}{\hbar^{2} v^{2}_{\scriptscriptstyle{\mathrm{F}}}} \frac{\pi \omega^{\0}_{\mathrm{K}}}{4}\!\! \int_{\vartheta}^{\pi}{\mathrm{d}\theta},
\end{equation}
with $\vartheta$ being the angle at which the shifted Fermi circle and the non-shifted circle of radius $\frac{\tilde{\omega}_{\mathrm{K}}}{2}$ intersect, i.e.,
\begin{equation}\label{VARPHI_SOLUTION}
\tilde{k}_{\scriptscriptstyle{\mathrm{F}}}(\vartheta) \cong 1 + \eta \cos{\vartheta} =  \tilde{\omega}_{\mathrm{K}}/2.
\end{equation}
Solutions for $\vartheta$ exists only if $\left||\tilde{\omega}_{\mathrm{K}}/2| - 1\right| \leq \eta$. The explicit solutions to Eqs.~(\ref{BROADENING_SIMPLIFIED_K_POINT}) and ~(\ref{VARPHI_SOLUTION}) are presented in Sec.~\ref{subsec:ARFACGSATEZ_K} by Eqs.~(\ref{ANALYTIC_FORMULA_FOR_MODE_BROADENING_FINALIZED_K})--(\ref{EF_PM_K}) in terms of $\psi = \pi - \vartheta$.
%%%%%%%%%%%%%%%%%%%%%%%%%%%%%%%%%%%%%%%%%%%%%%%%%%%%%%%%%%%
%BIBLIOGRAPHY
%%%%%%%%%%%%%%%%%%%%%%%%%%%%%%%%%%%%%%%%%%%%%%%%%%%%%%%%%%%
\clearpage
\bibliography{Manuscript_BiB}
\end{document}